\documentclass[final,3p,times]{elsarticle}

\usepackage{amssymb}
\usepackage{amsfonts}
\usepackage{textcomp, gensymb}
\usepackage{physics}
\usepackage{graphicx}
\usepackage{float}
\usepackage{tabularx}
\usepackage{adjustbox}
\usepackage{multirow}
\usepackage{xcolor}
\usepackage[%
    colorlinks=true,
    pdfborder={0 0 0},
    linkcolor=red
]{hyperref}
\usepackage{cleveref}

\journal{Energy and Buildings}
\begin{document}

\begin{frontmatter}

\title{Empirical exploration of zone-by-zone energy flexibility: a non-intrusive load disaggregation approach for commercial buildings}

\author[1]{Maomao Hu\corref{cor1}}
\author[2,3]{Ram Rajagopal}
\author[1]{Jacques A. de Chalendar}
\cortext[cor1]{Corresponding author}

\address[1]{Department of Energy Science \& Engineering, Stanford University, Stanford, CA, 94305, USA}
\address[2]{Department of Civil \& Environmental Engineering, Stanford University, Stanford, CA, 94305, USA}
\address[3]{Department of Electrical Engineering, Stanford University, Stanford, CA, 94305, USA}
            
\begin{abstract}
Building energy flexibility has been increasingly demonstrated as a cost-effective solution to respond to the needs of energy networks, including electric grids and district cooling and heating systems, improving the integration of intermittent renewable energy sources. Adjusting zonal temperature set-points is one of the most promising measures to unlock the energy flexibility potential of central air conditioning systems in complex commercial buildings. However, most existing studies focused on quantifying the energy flexibility on the building level since only building-level energy consumption is normally metered in today’s commercial buildings. By better understanding zone-level energy flexibility, building managers can design more targeted energy flexibility strategies, balancing overall flexibility potential with costs to occupants. This study aims to investigate the impacts of temperature set-point adjustment strategies on zone-level thermal and energy performance by developing a non-intrusive data-driven load disaggregation method (i.e., a ‘virtual’ zonal power meter). Three university buildings in Northern California (containing 136, 217, and 142 zones) were selected to test the proposed load disaggregation method. Zonal temperature set-point adjustment strategies were previously tested in these buildings. We found that heterogeneities of energy use and energy flexibility existed across not only buildings but also air handling units (AHUs) and zones. Moreover, a small number of zones accounted for a large amount of energy use and energy flexibility; and the most energy-intensive zones are not necessarily the most energy-flexible zones. For the three tested buildings, the top 30\% most energy-intensive zones accounted for around 60\% of the total energy use; and the top 30\% most energy-flexible zones provided around 80\% of the total energy flexibility. We also found the effects of temperature set-point adjustment on indoor air temperature were limited and heterogeneous. The proposed virtual zonal power meter enables the electric grid or district energy system operators to regard the controlled energy-flexible entities as a fleet of AHUs or zones instead of a fleet of buildings and helps unlock the possibility for targeted demand flexibility strategies that balance zone-by-zone energy reduction with zone-by-zone costs to occupants. 

\end{abstract}

\begin{keyword}
Energy flexibility \sep Load disaggregation \sep Thermal demand response
\end{keyword}

\end{frontmatter}

\section{Introduction}
Displacing conventional fossil fuels with renewable energy sources (RESs) plays a significant role in the decarbonization of today’s electric grids. However, due to the intrinsic intermittency and uncertainty of the generation of RESs, electric grids are facing an increasing challenge in managing the balance between the supply and demand sides \cite{Niknam2012-kl, Neves2016-ga}. Buildings account for around 40\% of the total energy consumption worldwide and approximately 30\% of global greenhouse gas emissions \cite{Iea2019-xm}. In the United States, the buildings sector is responsible for around 75\% of the total electricity consumption and contributes to about 80\% of the peak demand \cite{Center2020-hq, Eckman2022-qo}. Buildings have demonstrated capabilities of providing energy flexibility to alleviate the power imbalance issue and integrate more RESs into electric grids \cite{Hu2021-yw, Li2021-ox, Li2022-gs}.

The major energy flexibility measures for buildings include the integration of on-site distributed power generation, the utilization of thermal or battery storage systems, and the adjustment of schedulable electrical equipment \cite{Hu2021-yw}. Heating, ventilation, and air conditioning (HVAC) systems are one of the most promising schedulable resources in buildings to reduce or shift electricity use during peak hours \cite{Wang2019-kt, Watson2006-cp}. This is because HVAC systems account for around 50\% of the total energy used in buildings in the US \cite{Li2012-lu} and buildings, especially commercial buildings, are equipped with advanced building automation and management systems.

Temperature set-point adjustments are one of the most common energy flexibility measures for HVAC systems in buildings. The implementation of temperature set-point adjustments can leverage the existing building management system and does not require significant reprogramming work. In contrast to the direct control of HVAC equipment such as chillers, fans, and pumps \cite{Tang2018-jw, Wang2020-di}, temperature set-point adjustments have been demonstrated to be an easily implementable, scalable, and low-cost method to provide energy flexibility \cite{De_Chalendar2023-zh}. Many studies, including simulation-based and experimental studies, have focused on evaluating the energy flexibility potential of temperature set-point adjustment strategies. Yin et al. \cite{Yin2016-iy} investigated the hourly power reduction potential of the cooling systems in both residential and commercial buildings by adopting temperature set-point adjustment strategies. Based on exhaustive simulations using EnergyPlus \cite{Crawley2000-ji}, a simplified regression model was developed to estimate the energy flexibility potential based on various inputs such as set-point change and outdoor air temperature. Hu et al. \cite{Hu2017-tz, Hu2018-dg} adopted temperature set-point adjustments combined with pre-cooling strategies to optimally shift the energy consumption of residential air conditioners from high-price to low-price periods with the assistance of a model-based optimal control method. Simulation results showed that optimal rescheduling of temperature set points can help reduce the peak power demands during peak hours while still maintaining the thermal comfort of occupants in the dynamic electricity pricing environment. Regarding experimental studies, Kiliccote et al. concluded that commercial buildings in California can provide 13\% reductions in peak power demands on average (5\% - 15\%) by mostly using the global temperature adjustment strategy \cite{Kiliccote2010-re}. The thermal comfort of occupants needs to be considered when implementing any type of energy flexibility measures. Simulation-based and field studies in the literature suggested that the temperature set-point adjustment strategies did not have significant negative impacts on occupants’ thermal comfort when the temperature set-point was less than 28\degree C \cite{Aghniaey2018-ms}. 

Although there have been extensive studies on the assessment of the energy flexibility potential of temperature set-point adjustment strategies in multi-zone commercial buildings, most existing studies focused on quantifying the energy flexibility on the building level since only building-level energy consumption is normally measured in today’s commercial air-conditioned buildings. However, understanding zone-level energy flexibility with high granularity is desirable, so building managers can selectively apply energy flexibility measures to the most energy-flexible zones while considering the thermal comfort of occupants. To achieve this goal, an effective approach to disaggregating the building-level energy consumption into zone-level load demands is needed.

To bridge the research gap in the literature, we investigated the impacts of temperature set-point adjustment strategies on zonal thermal and energy performance by developing a data-driven non-intrusive load disaggregation method. The key idea of the non-intrusive load disaggregation method is to develop a ‘virtual’ zonal power meter to provide zone-level energy performance. We tested the proposed load disaggregation method on three university buildings in Northern California (containing 136, 217, and 142 zones) where zone-level temperature strategies were previously tested \cite{De_Chalendar2023-zh}. Specifically, the major contributions of our study are as follows. First, we developed a data-driven non-intrusive load disaggregation method and test results show that it can effectively disaggregate the metered building-level cooling load into the zonal equivalent electrical load considering the fan power consumption. The models used in the method are physically interpretable, simple, and scalable. With the wide availability of IoT sensors in modern commercial buildings, the load disaggregation technique can help provide building owners with high-granular insights into zonal energy performance, including energy use and energy flexibility. Second, heterogeneities of energy use and energy flexibility were found across not only buildings but also air handling units (AHUs) and zones. AHUs and zones in a building did not identically respond to the cooling temperature set-point adjustment. We also found a small number of zones accounted for a large amount of energy use and energy flexibility. For the three tested buildings, the top 30\% of most energy-intensive zones account for around 60\% of the total energy use; and the top 30\% of most energy-flexible zones provide around 80\% of the total energy flexibility. With these findings, the operators of electric grids and district energy systems can regard the controlled DR entities as a fleet of AHUs or zones instead of a fleet of buildings, which helps unlock the possibility for targeted demand flexibility strategies that balance zone-by-zone energy reduction with zone-by-zone costs to occupants. Third, we found that indoor air temperatures and the degrees of over-cooling (i.e., energy is wasted on excessive cooling of rooms) were heterogeneous across zones. Moreover, the impacts of temperature set-point adjustment on indoor air temperature were found to be limited and heterogeneous. Therefore, zone-level thermal dynamics under certain temperature set-point adjustments and outdoor air conditions need to be predicted before rolling out energy flexibility measures in practice.

\section{Experimental testbed and design}
\subsection{Target buildings and HVAC systems}
Three campus buildings in Northern California were selected as the target buildings to test the proposed load disaggregation approach. Daily temperature set-point adjustment strategies were previously tested on these buildings \cite{De_Chalendar2023-zh}. We provide a brief description of the experiment test bed, data collection systems, and experiment protocols here. More details can be found in our previous publication \cite{De_Chalendar2023-zh}.
\par
\begin{figure}[ht]
    \centering
    \includegraphics[width=1\textwidth]{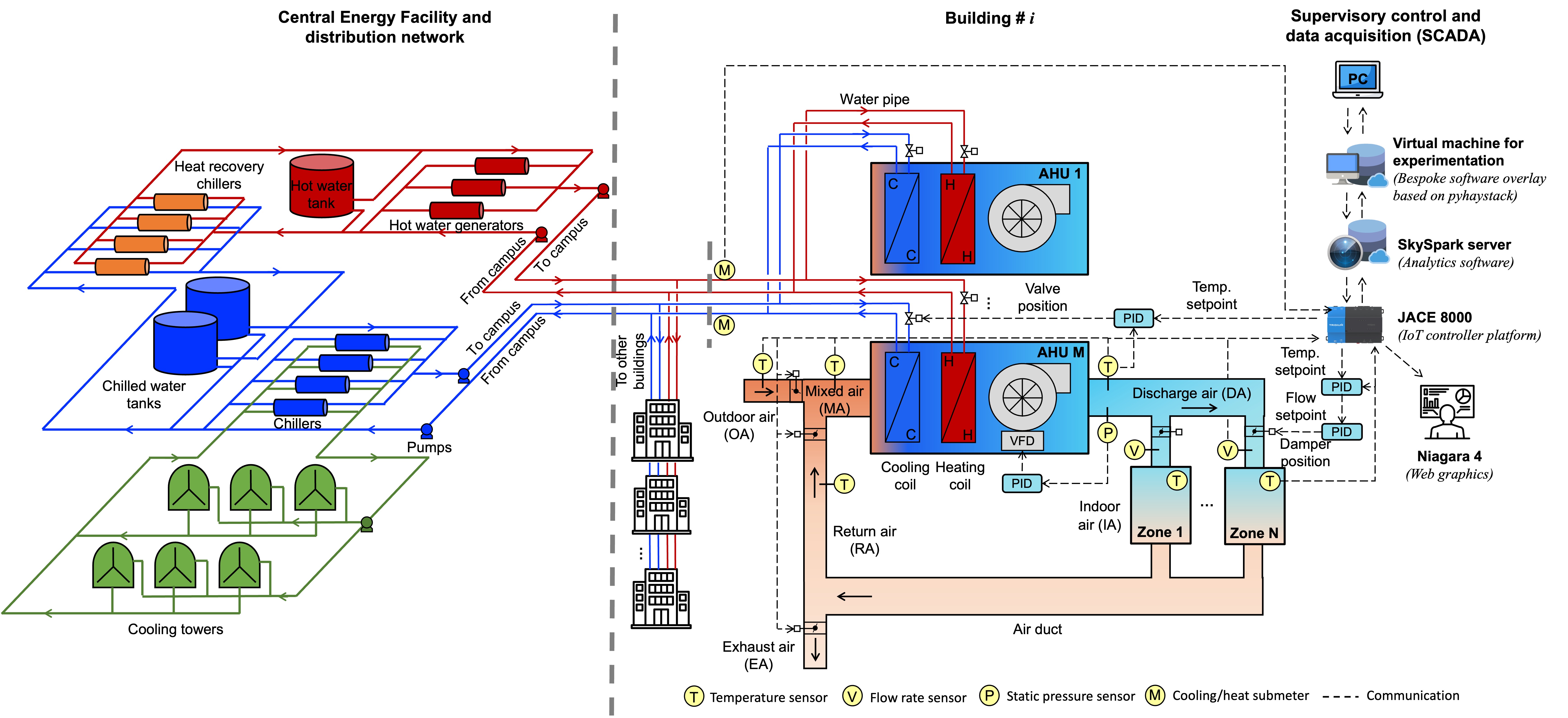}
    \caption{Schematic diagram of the target complex cyber-physical district energy system}
    \label{fig: system diagram}
\end{figure}

 The main characteristics of the target buildings are shown in \Cref{tab:characteristics_of_buildings}. As shown in \Cref{fig: system diagram}, cooling in the target buildings is supplied by a Central Energy Facility through a chilled water loop system across the campus. Inside the building, a central air-conditioning system, consisting of AHUs and variable air volume (VAV) boxes, is used to modulate the cooling supply to a building. The basic control principle of the central air-conditioning system is as follows: the opening degree of the chilled water valve is modulated to maintain the supply air temperature at the pre-defined set-point. The motor of the supply fan is based on pressure control to maintain the static pressure set-point to distribute sufficient air flow to each room. At the room side, the damper of the VAV terminals is controlled to maintain the measured indoor air temperature between the cooling and heating set points, the ``deadband''.
\begin{table}[ht!]
    \centering
    \caption{
    \label{tab:characteristics_of_buildings} Main characteristics of target experimental buildings}
    \begin{tabular}{llll}
    \hline \hline
     & Bldg. A & Bldg. B & Bldg. C \\
    \hline
    Type & Office/Conference  &  Office/Classroom & Office/Classroom \\
    Year of Construction & 2000 & 1998 & 1996 \\
    Year of last retrofit & 2021 & / & 2021\\  
    \# of AHUs & 5 & 4 & 9\\
    \# of VAVs & 136 & 217 & 142 \\
    Floor area (1,000 m$^2$) & 13.5 & 9.8 & 15.8\\ 
    \hline 
    \hline
    \end{tabular}%
\end{table}

\subsection{Experimentation software systems and data collection}
A bespoke software overlay was developed to automatically send commands to and collect data from the pre-existing building energy management systems. It was written in Python and leverages functionality from the open-source pyhaystack module \cite{Tremblay2022-wc}, which enables users to communicate with a server based on Haystack semantic model. Haystack is an open-source project to standardize semantic models for processing IoT data \cite{Project_Haystack2022-ti}. In the target buildings, the Haystack server (Skyspark server) is used to communicate with network controllers (JACE 8000, typically one per floor), which connect to zone-level controllers. The software overlay can also connect to a separate data historian to retrieve historical data of chilled water usage per building. Outdoor air dry-bulb temperature can be continuously recorded from the campus weather station. \Cref{tab: measured variables} summarizes the measured variables in the experimental testbed.

\begin{table}[ht!]
    \centering
    \caption{
    \label{tab: measured variables} Measured variables in the experimental testbed at different levels}
    \begin{tabular}{llll}
    \hline \hline
     Measurement level & Variable & Symbol & Unit \\
    \hline
    District level & Total electrical load for cooling generation at time $t$  & $p_{d,t}$ & kW \\
     & Total supplied cooling load at time $t$ & $q_{d,t}$ & kW \\
    \hline
    Building level & Building-level cooling load at time $t$ & $q_{b,t}$ & kW\\  
     & Global indoor air temperature set point at time $t$ & $SP_t$ & \degree C \\ 
    \hline
    AHU level & Return air temperature for AHU $i$ at time $t$ & $RAT_t^i$ & \degree C \\ 
     & Mixed air temperature for AHU $i$ at time $t$ & $MAT_t^i$ & \degree C \\
     & Discharge air temperature for AHU $i$ at time $t$ & $DAT_t^i$ & \degree C \\
    \hline
    Zone level & Indoor air temperature for Zone $j$ in the AHU $i$ at time $t$ & $IAT_t^{i,j}$ & \degree C \\
     & Zone-level air volumetric flow rate for Zone $j$ in the AHU $i$ at time $t$ & $v_{z,t}^{i,j}$ & $m^3/s$ \\ 
     \hline
    Weather condition & Outdoor air dry-bulb temperature at time $t$ & $OAT_t$ & \degree C \\
    \hline 
    \hline
    \end{tabular}%
\end{table}

\subsection{Experimental design}
The experiments took place from June 22 to August 15, 2021 (54 days in total). During the experiments, cooling setpoints of zone-level indoor air temperature were adjusted every two days at 6 AM and periodically changed between a low value (23.3\degree C, 74\degree F, baseline case) and a high value (24.4\degree C, 76\degree F, DR case). The same daily setpoint command was broadcast to all zones in the three experimental buildings on each day. In every building, a subset of sensitive zones was excluded from the experiments.

\section{Load disaggregation method}
\subsection{Problem statement and overview of the proposed method}

\begin{figure}[ht]
    \centering
    \includegraphics[width=0.7\textwidth]{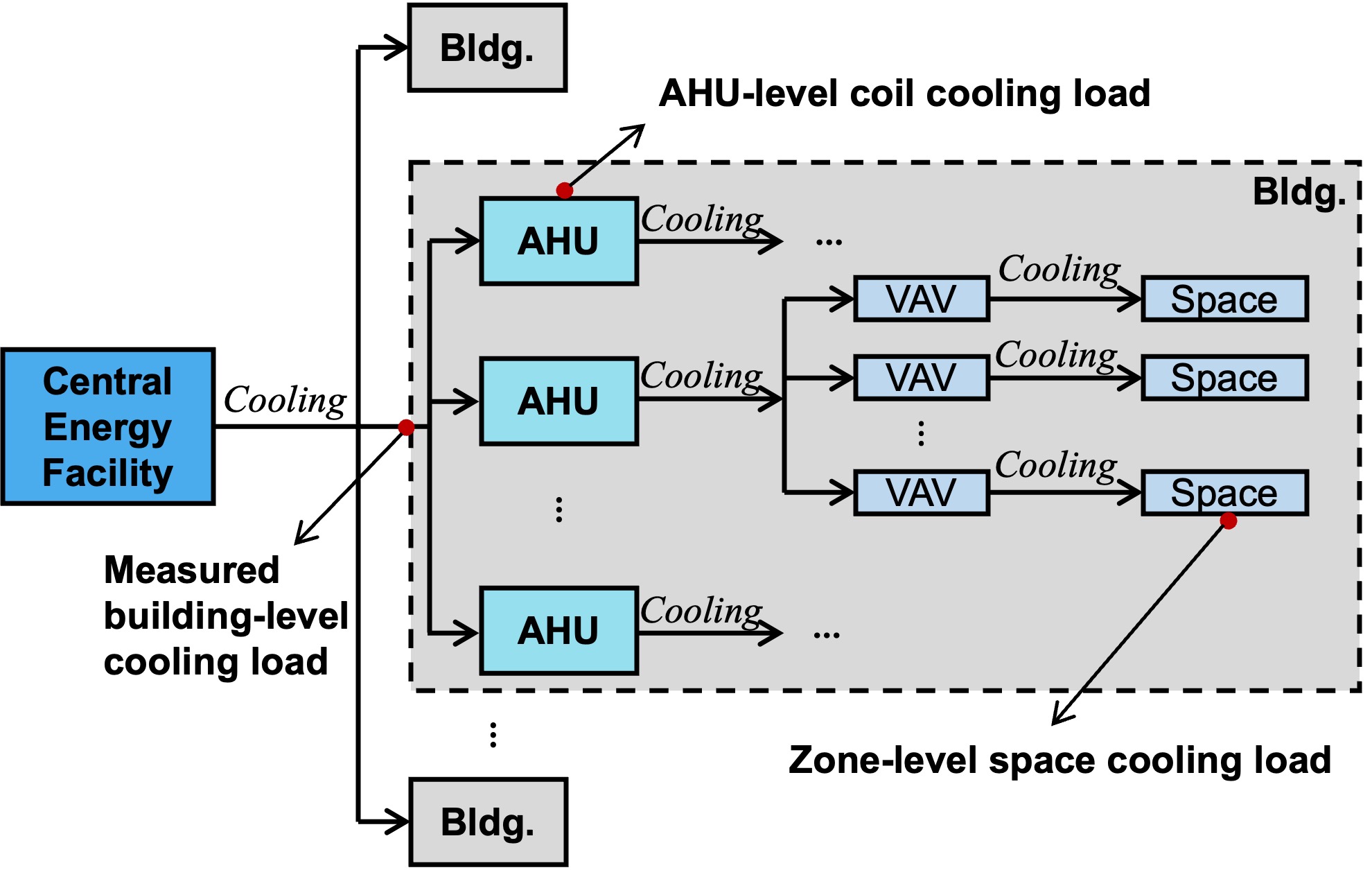}
    \caption{Schematic diagram of the cooling load flow across various levels from building through AHUs to zones}
    \label{fig: cooling load flow}
\end{figure}

Unlike our previous study \cite{De_Chalendar2023-zh} in which we evaluated building-level flexibility in cooling load, the goal of this study is to investigate the high-granular AHU-level and zone-level flexibility in electrical load with the consideration of both cooling load and the power consumption of supply air fans. Regarding cooling load, \Cref{fig: cooling load flow} shows the three different cooling load levels we consider: for buildings, AHUs, and zones. Only the building-level cooling load is measured by a dedicated meter, which was used to evaluate the building-level flexibility in cooling load in \cite{De_Chalendar2023-zh}. For AHU- and zone-level cooling loads, they can be indirectly calculated using physical variables (listed in \Cref{tab: measured variables}) measured by different sensors and \Crefrange{eq: coil cooling load 1}{eq: single zone space air cooling load}. However, we can neither use the coil cooling load (\Cref{eq: coil cooling load 1}) as the AHU-level energy consumption nor use the space cooling load (\Cref{eq: single zone space air cooling load}) as the zone-level energy consumption in practice. This is due to the issue of mismatched cooling loads at different levels we identified when using real-life experimental data in target buildings. Therefore, the main research question we try to answer here is how we can develop physically interpretable regression models to characterize the gaps between the mismatched cooling loads at different levels and how we can then disaggregate the cooling load gaps into AHU and zone levels. In addition to the cooling load consumed in the cooling coils, the building air conditioning systems use electrical energy to power the supply air fans. Since the supply air fans are not sub-metered in the testbed, energy performance models of supply fans are also needed to estimate the power consumption of fans. The illustration of the mismatched cooling load issue using on-site data and a brief overview of the proposed method are presented below.

\begin{figure}[ht]
    \centering
    \includegraphics[width=1\textwidth]{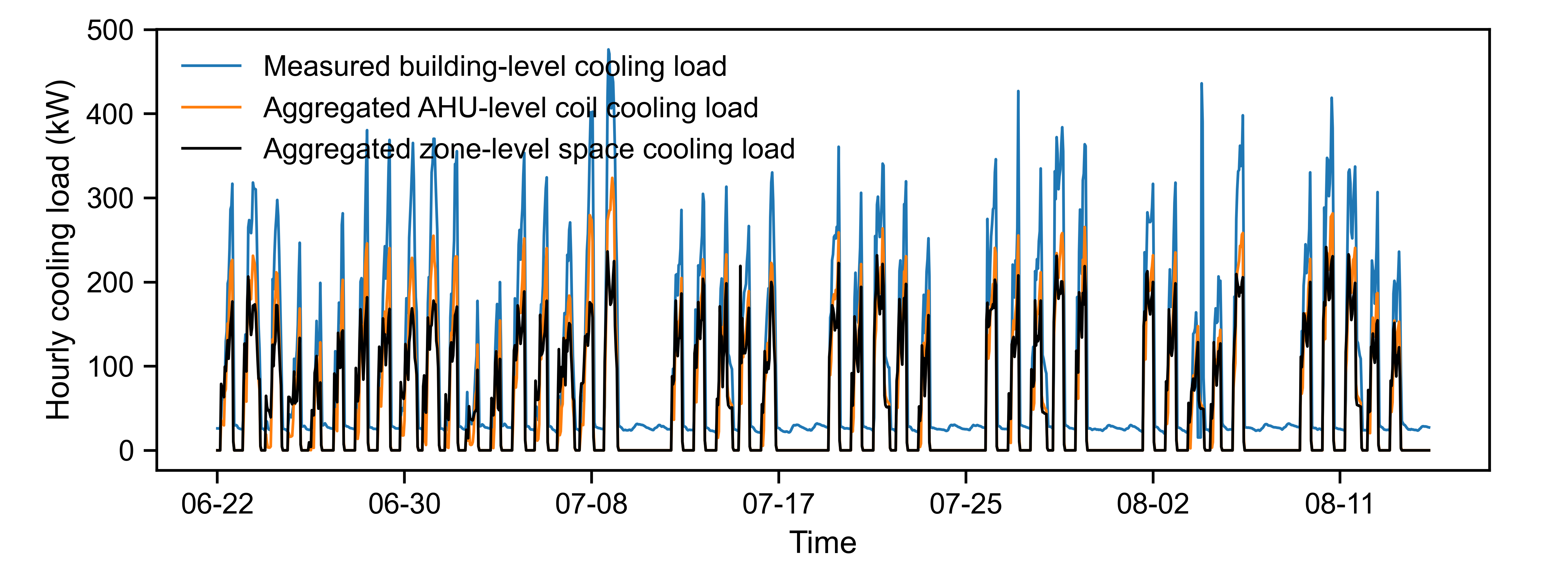}
    \caption{Illustrative example of the issue of the mismatched cooling loads at different levels in one target building (Bldg. A).}
    \label{fig: different levels of cooling loads}
\end{figure}

\paragraph{Illustration of the mismatched cooling load issue using on-site data} We call $q_{b,t}$ the cooling load of a target building at time $t$. We define the coil cooling load of AHU $i$ at time $t$ as
\begin{equation}
\label{eq: coil cooling load 1}
    q_{c,t}^i = c\rho v_{c,t}^i(MAT_t^i - DAT_t^i),
\end{equation}
where $DAT_{t}^{i}$ and $MAT_t^i$ are the measured discharge and ``mixed'' (entry) temperatures for AHU $i$, respectively, $v_{c,t}^i=\sum_{j=1}^N v_{z,t}^{i,j}$ is the sum of the volumetric flow rates of air measured at the entrance to each zone $j$ served by AHU $i$, and $c$ and $\rho$ denote the specific heat capacity and density of air. We define the space air cooling load of zone $j$ served by AHU $i$ at time $t$ as
\begin{equation} 
\label{eq: single zone space air cooling load}
    q_{z,t}^{i,j} = c\rho v_{z,t}^{i,j}(IAT_{z,t}^{i,j}-DAT_{t}^{i}),
\end{equation}
where we additionally use measurements for indoor air temperature of zone $j$, $IAT_{z,t}^{i,j}$. We use the subscripts $c$ and $z$ to denote cooling coils and zones, respectively.

Illustrative data are shown in \Cref{fig: different levels of cooling loads} for one of the measured cooling load of target buildings $q_{b,t}$ (blue), the same building's aggregate AHU cooling load $\sum_{i=1}^M q_{c,t}^i$ (orange), and the building's aggregate space cooling load $\sum_{i=1}^M \sum_{j=1}^N q_{z,t}^{i,j}$ (black). The mismatch between these three lines can partly be assigned a physical interpretation and partly be explained by measurement error. We develop data-driven regression models in the next section to fill the two gaps, from $\sum_{i=1}^M \sum_{j=1}^N q_{z,t}^{i,j}$ to $\sum_{i=1}^M q_{c,t}^i$ and from $\sum_{i=1}^M q_{c,t}^i$ to $q_{b,t}$ based on historical data and modeling assumptions.

\begin{figure}[ht]
    \centering
    \includegraphics[width=1\textwidth]{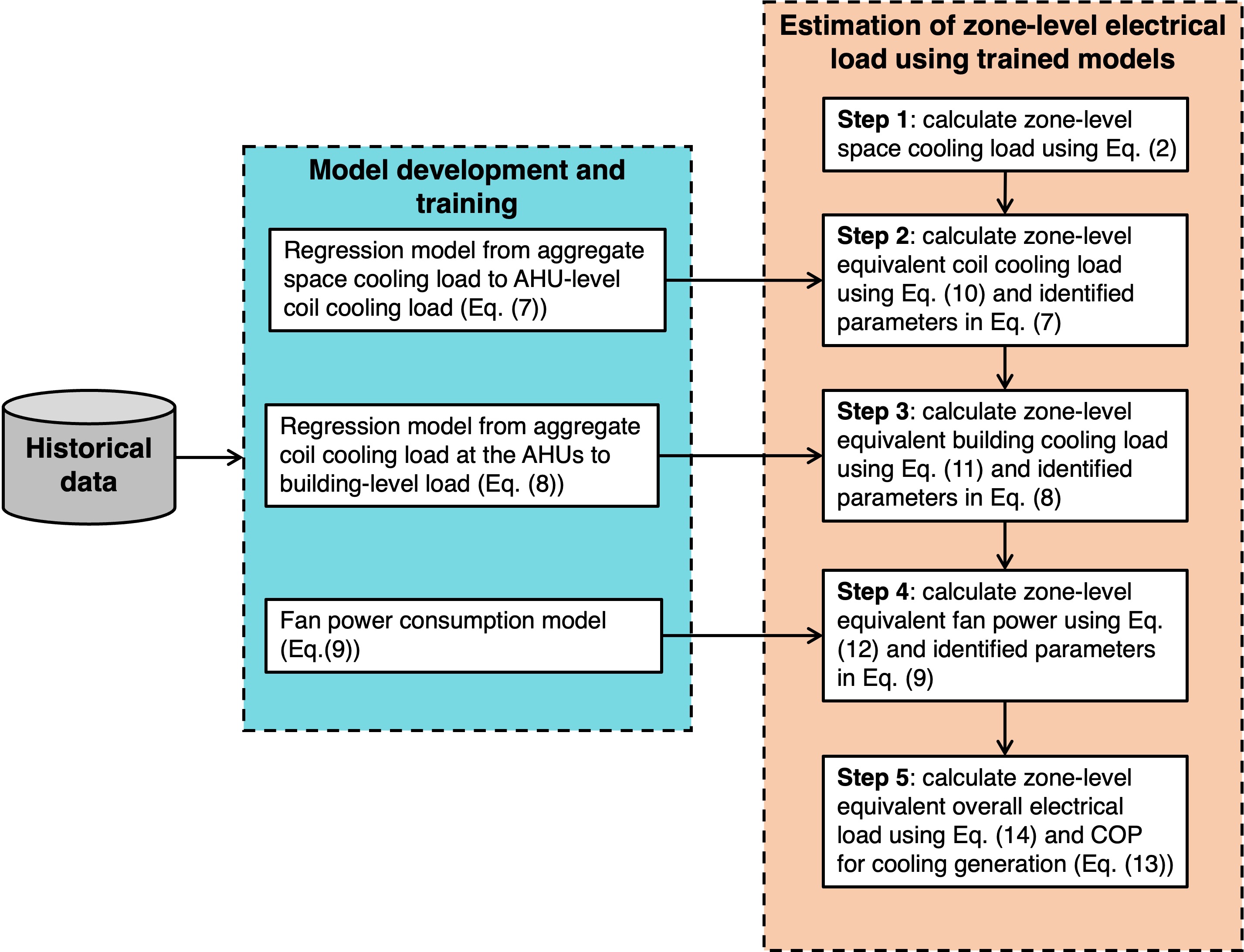}
    \caption{Flowchart of the proposed load disaggregation method.}
    \label{fig: flowchart of the proposed load disaggregation method}
\end{figure}

\paragraph{Overview of the load disaggregation method}
As shown in \Cref{fig: flowchart of the proposed load disaggregation method}, the proposed data-driven load disaggregation approach consists of two stages.  In Stage 1 (\Cref{sec: development of regression models}), we aim to develop and train the regression models to fill the gaps between the cooling loads at different levels (\Crefrange{eq: fresh air load regression model}{eq: aggregate coil cooling}) and the fan power consumption model (\Cref{eq: fan power}). After the development of the models, we identify the unknown parameters in the models using historical data. In Stage 2 (\Cref{sec: forward estimation}), we estimate the zone-level electrical load based on the space cooling load (\Cref{eq: single zone space air cooling load}) and trained models. The essence of the estimation of the zone-level electrical load is adjusting the zone-level space cooling load by disaggregating the cooling load gaps between different levels.

\subsection{Development of regression models}
\label{sec: development of regression models}

\subsubsection{From aggregate space cooling load to AHU-level coil cooling load} 
To relate the aggregate space cooling load to the AHU-level coil cooling load, we start by introducing the return air temperature of AHU $i$
\begin{equation}
    RAT_t^i=\frac{\sum_{j=1}^N v_{z,t}^{i,j}IAT_{z,t}^{i,j}}{v_{c,t}^i},
\end{equation}
which we can use to write the aggregate space cooling load for AHU $i$ as
\begin{equation}
    \sum_{j=1}^{N}q_{z,t}^{i,j} = \sum_{j=1}^{N} c\rho v_{z,t}^{i,j}(IAT_{z,t}^{i,j}-DAT_{t}^{i}) = c\rho v_{c,t}^i (RAT_t^i- DAT_{t}^{i}).
\end{equation}

Separately, we can write the mixed air temperature for AHU $i$ as the weighted sum of $RAT_t^i$ and outside air temperature $OAT_t^i$,
\begin{equation}
    MAT_t^i = k^i OAT_t^i + (1-k^i)RAT_t^i,
\end{equation}
where we assumed that the fresh air ratio $k^i$ is independent of time. From the definition of coil cooling load in \Cref{eq: coil cooling load 1}, we have
\begin{equation}
\begin{split}
    q_{c,t}^i &= c\rho v_{c,t}^i(MAT_t^i - DAT_t^i),\\
    &= c\rho v_{c,t}^i\left(k^i (OAT_t^i-RAT_t^i) +(RAT_t^i - DAT_t^i)\right),\\
    &= c\rho v_{c,t}^ik^i (OAT_t^i-RAT_t^i) + \sum_{j=1}^{N}q_{z,t}^{i,j}. 
\end{split}
\label{eq:fresh air load}
\end{equation}
We call $c\rho v_{c,t}^ik^i (OAT_t^i-RAT_t^i)$ the ``fresh air'' load. The cooling coils in the AHUs are responsible for cooling the mixture of return air and outdoor air to the discharge air temperature set point. From \Cref{eq:fresh air load}, the cooling coil load can be divided into space cooling load and fresh air load. Fresh air load increases proportionally to the difference between the outdoor and return air temperatures. 

The mismatch between the coil cooling load and aggregate space cooling load can partly be explained by the fresh air load and partly by measurement error. We estimate a regression model for the fresh air load of AHU $i$ of the form
\begin{equation} 
\label{eq: fresh air load regression model}
\frac{q_{c,t}^{i} - \sum_{j=1}^{N}q_{z,t}^{i,j}}{c\rho v^i_{c,t}} = k^i(OAT_{t} - RAT_{t}^{i}) + \alpha^i + \epsilon^i_t.
\end{equation}
In \Cref{eq: fresh air load regression model}, $k^i$ and $\alpha^i$ are parameters to be estimated, and $\epsilon^i_t$ is a random error term with mean zero. We expect $0 \leq k^i\leq 1$ as $k^i$ represents the fresh air ratio, which is the ratio of the outdoor to total supply air flow rates. $\alpha^{i}$ can be interpreted as the average energy loss through the air distribution system.

\subsubsection{From aggregate coil cooling load at the AHUs to building-level load} We similarly use a regression model for the gap we observe from the aggregate coil cooling load ($\sum_{i=1}^{M}q_{c,t}^{i}$) to the building-level load ($q_{b,t}$):
\begin{equation}
    \label{eq: aggregate coil cooling} 
    q_{b,t} = l^b\sum_{i=1}^{M}q_{c,t}^{i} + \beta^b +\epsilon^b_t,
\end{equation}
where $l^b$ and $\beta^b$ are the parameters to be identified by fitting recorded data for a target building and $\epsilon^b_t$ is a random error term with mean zero. The mismatch between aggregate coil cooling load and building-level load can partly be explained by uses of chilled water outside of the AHUs, e.g. for water–cooled equipment, losses in the chilled water system, and measurement errors.

\subsubsection{Electricity consumption of the fans} According to Ref. \cite{Brelih2012-pq}, fan accounts for around 40\% of the total energy consumed by the central air-conditioning system. In our study, the fan power consumption for AHU $i$, $p^i_{fan,t}$, is approximated using a third-order polynomial function of the total supply air flow rate across the AHU \cite{Tang2018-ch, Ma2012-rc}, as shown in \Cref{eq: fan power}.
\begin{equation}
    \label{eq: fan power}
    p^i_{fan,t} = a^i_0 + a^i_1 v^i_{c,t} + a^i_2 (v^i_{c,t})^2 + a^i_3 (v^i_{c,t})^3 + \epsilon^i_t,
\end{equation}
where $a^i_0$, $a ^i_1$, $a^i_2$, and $a^i_3$ are the parameters to be identified by fitting recorded data for AHU $i$, and $\epsilon^i_t$ is a random error term with mean zero.

\subsection{Estimation of zone-level electrical load using regression models}
\label{sec: forward estimation}
After identifying the unknown parameters in \Crefrange{eq: fresh air load regression model}{eq: fan power}, the overall energy consumption of each zone, including cooling load and electrical load from the supply fan, can be estimated by using \Cref{eq: coil cooling load 1,eq: single zone space air cooling load,eq: zone-level equivalent coil cooling load,eq: zone-level equivalent building cooling load,eq: zone-level equivalent fan power,eq: district COP,eq: zone-level equivalent overall electrical load}. The key idea behind the estimation of zone-level electrical load is to divide up the gaps between different levels of cooling loads proportionally to the ratio of the zone-level flow rate to the AHU-level flow rate, as shown in \Cref{eq: zone-level equivalent coil cooling load,eq: zone-level equivalent building cooling load}.

\begin{enumerate}
    \item Zone-level equivalent coil cooling load at time $t$
    \begin{equation}
        \label{eq: zone-level equivalent coil cooling load}
        \hat{q}_{z,ec,t}^{i,j} = q_{z,t}^{i,j} + c\rho v_{z,t}^{i,j} (k^i(OAT_t - RAT_t^i) + \alpha^i)
    \end{equation}

    \item Zone-level equivalent building cooling load at time $t$
    \begin{equation}
        \label{eq: zone-level equivalent building cooling load}
        \hat{q}_{z,eb,t}^{i,j} = l^b\times \hat{q}_{z,ec,t}^{i,j} + \frac{v_{z,t}^{i,j}}{v^i_{c,t}}\times\frac{q_{c,t}^{i}}{\sum_{i=1}^M q_{c,t}^{i}}\times\beta^b
    \end{equation} 

    \item Zone-level equivalent fan power at time $t$
    \begin{equation}
        \label{eq: zone-level equivalent fan power}
        \hat{p}_{z,fan,t}^{i,j} = \frac{v_{z,t}^{i,j}}{v^i_{c,t}}\times p_{fan,t}^i
    \end{equation}

    \item District-level coefficient of performance ($COP$) for cooling generation at time $t$
    \begin{equation}
        \label{eq: district COP}
        COP_{d,t} = \frac{q_{d,t}}{p_{d,t}}
    \end{equation}
    
    \item Zone-level equivalent overall electrical load at time $t$
    \begin{equation}
        \label{eq: zone-level equivalent overall electrical load}
        \hat{p}_{z,t}^{i,j} = \frac{\hat{q}_{z,eb,t}^{i,j}}{COP_{d,t}} + \hat{p}_{z,fan,t}^{i,j}
    \end{equation}
\end{enumerate}

\section{Performance evaluation}
\subsection{Energy flexibility and energy flexibility share}
In our experimentation, the DR control strategy is to increase the indoor air temperature setpoint by 2\degree F (1.1\degree C). The energy flexibility ($EF$) and energy flexibility share ($EFS$) of a specific zone $j$ served by AHU $i$ can be quantified by using \Cref{eq: performance index - energy flexibility,eq: performance index - energy flexibility share}, respectively.

    \begin{equation}
        \label{eq: performance index - energy flexibility}
        EF_{z}^{i,j} = \frac{\Bar{E}_{z,LSP}^{i,j}-\Bar{E}_{z,HSP}^{i,j}}{\Bar{E}_{z,LSP}^{i,j}}
    \end{equation}
    
    \begin{equation}
        \label{eq: performance index - energy flexibility share}
        EFS_{z}^{i,j} = \frac{\max (\Bar{E}_{z,LSP}^{i,j}-\Bar{E}_{z,HSP}^{i,j},0)}{\sum_{i=1}^M \sum_{j=1}^N \max (\Bar{E}_{z,LSP}^{i,j}-\Bar{E}_{z,HSP}^{i,j},0)}
    \end{equation}

where $LSP$ and $HSP$ denote the days with low set-point and high set-point, respectively; $\Bar{E}$ denotes the average value of the daily energy consumption (kWh). Note that the energy flexibility of a specific zone ($EF_{z}^{i,j}$) could be negative based on real-life experimental data, which means the zone consumes more energy even though the indoor air cooling set-point is increased by 2\degree F in our case. On the contrary, the energy flexibility share is greater than or equal to zero, and the sum of the energy flexibility share of all zones in a building equals one, i.e., $EFS_{z}^{i,j} \geq 0$ and $\sum_{i=1}^M \sum_{j=1}^N EFS_{z}^{i,j} =1$.

\subsection{Quantification of heterogeneity of multi-zone energy use and energy flexibility}
\begin{figure}[ht]
    \centering
    \includegraphics[width=0.5\textwidth]{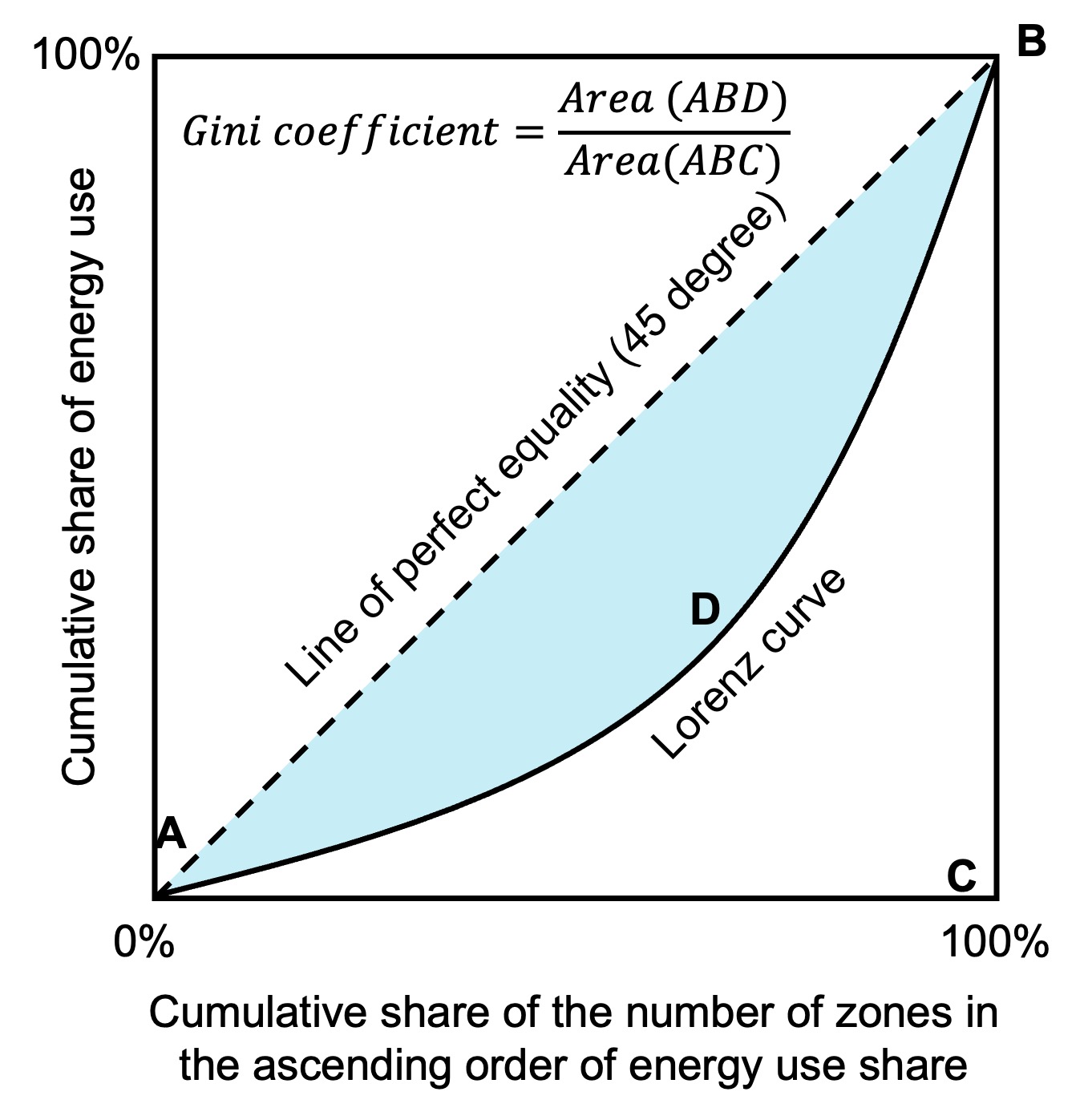}
    \caption{Lorenz curve and Gini coefficient for the quantification of heterogeneity of multi-zone energy use (adapted from Fig. 3 in \cite{Zhou2016-wy})}
    \label{fig:lorenz and gini}
\end{figure}

For a better understanding of the heterogeneous demand behavior across zones in a building, we use the Lorenz curve and Gini coefficient, which are commonly used in economics to describe income inequality, to quantitatively describe the heterogeneity of energy use and energy flexibility across zones \cite{Zhuang2021-dl,Zhou2016-wy}. \Cref{fig:lorenz and gini} shows the Lorenz curve for energy use distribution and the definition of the Gini coefficient. The X-axis represents the cumulative share of the number of zones in the ascending order of the energy use share. The Y-axis represents the cumulative share of energy use. The 45\degree\ line (Line A-B) represents perfect equality in energy use among the zones. When the Lorenz curve (Line A-D-B) greatly deviates from Line A-B, the energy demands differ significantly among the zones. Based on the Lorenz curve, the heterogeneity of energy use among zones can be quantified by using the Gini coefficient, which is the ratio of the area between the line of equality and the Lorenz curve (Area A-B-D) to the total area under the line of equality (Area A-B-C). The Gini coefficient can also be mathematically expressed as

\begin{equation}
    \label{eq: performance index - gini index}
    Gini = 1-\frac{1}{n^2\Bar{y}}\sum_{i=1}^{n}(2n-2i+1)y_i,
\end{equation}

where $n$ is the number of zones; $y$ denotes the energy use share of each zone; and $\Bar{y}$ denotes the average value of the energy use share. Note that the Gini coefficient for the energy use distribution is in the range of $[0,1]$, and a smaller value of the Gini coefficient indicates a higher similarity in energy demand behavior across zones. In our study, the Lorenz curve and Gini coefficient are also used to describe the heterogeneity of energy flexibility among zones.

\newpage
\section{Results}

\subsection{Model identification and validation}
\begin{figure}[ht]
    \centering
    \includegraphics[width=1\textwidth]{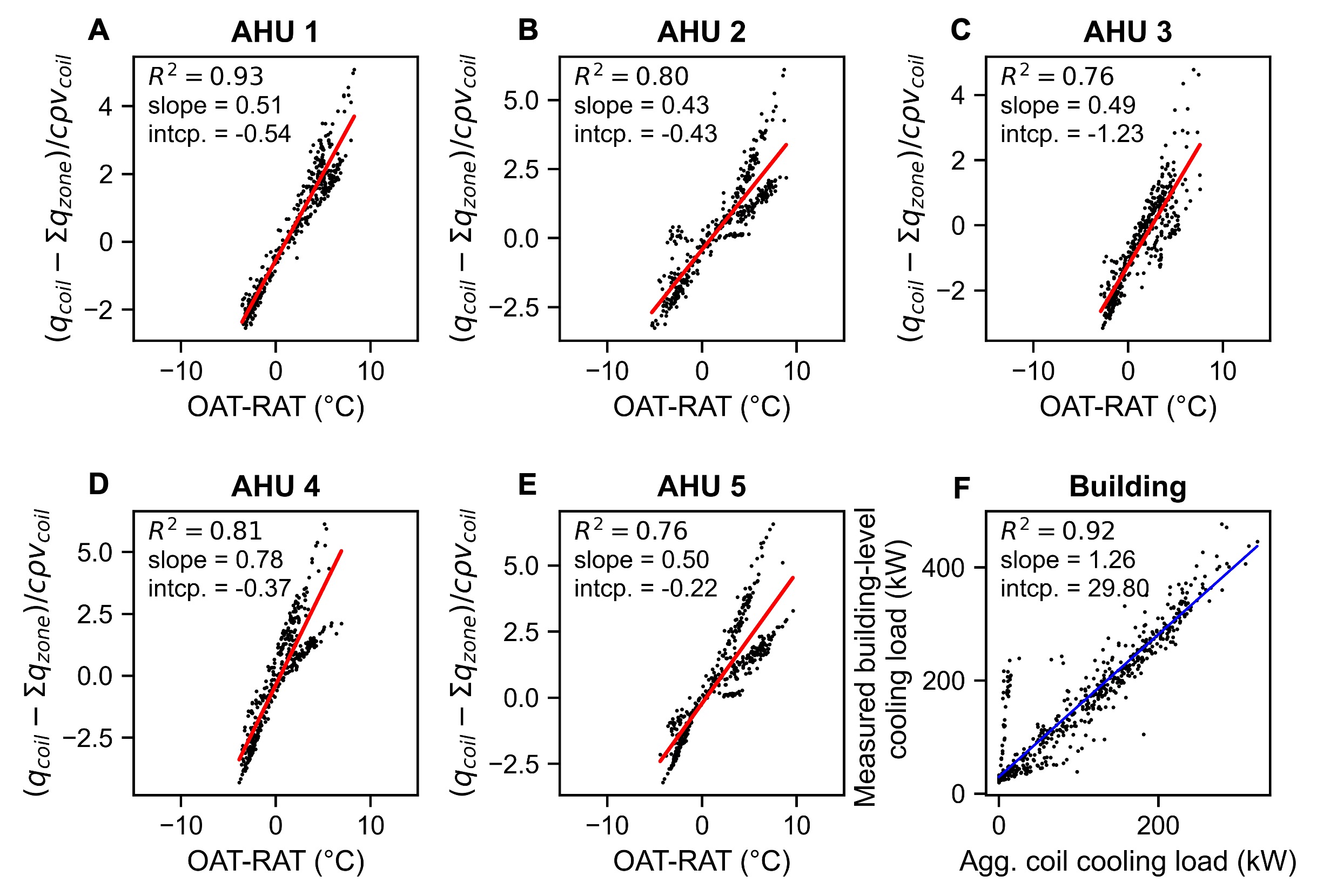}
    \caption{Identification results for Bldg. A during training session. (A-E) Regression from aggregate zone-level space cooling load to AHU coil cooling load for each AHU; (F) Regression from aggregate coil cooling load to measured building cooling load}
    \label{fig:regression models_bldg_A}
\end{figure}

The key step of load disaggregation is to develop regression models to fill the gap between aggregate zone-level space cooling load and AHU coil cooling load (Coil-Spaces model, \Cref{eq: fresh air load regression model}), and the gap between aggregate coil cooling load and measured building cooling load (Building-Coils model, \Cref{eq: aggregate coil cooling}). The entire dataset for each target building (54 days) is divided into training and validation data with proportions of 80\% (43 days) and 20\% (11 days).\Cref{fig:regression models_bldg_A} A-E show the regression results of the Coil-Spaces model for each AHU in Bldg. A using training data and \Cref{fig:regression models_bldg_A}-F shows the regression result of the Building-Coils model using training data. The coefficient of determination, $R^2$, ranges from 0.76 to 0.93, which indicates the regression models can accurately estimate the gaps between different levels of cooling loads. Note that only the operational points during the daytime on weekdays (i.e., when AHU systems are turned on) are included in the regression for AHUs. Additional numerical data for the regression models of Bldg. A during the training session and regression results for Bldg. B and Bldg. C can be found in \ref{sec: appendix}. The performance of regression models during both training and validation sessions are listed in \Cref{tab: comparison between training and validation sessions} for comparison.

\begin{figure}[ht]
    \centering
    \includegraphics[width=0.6\textwidth]{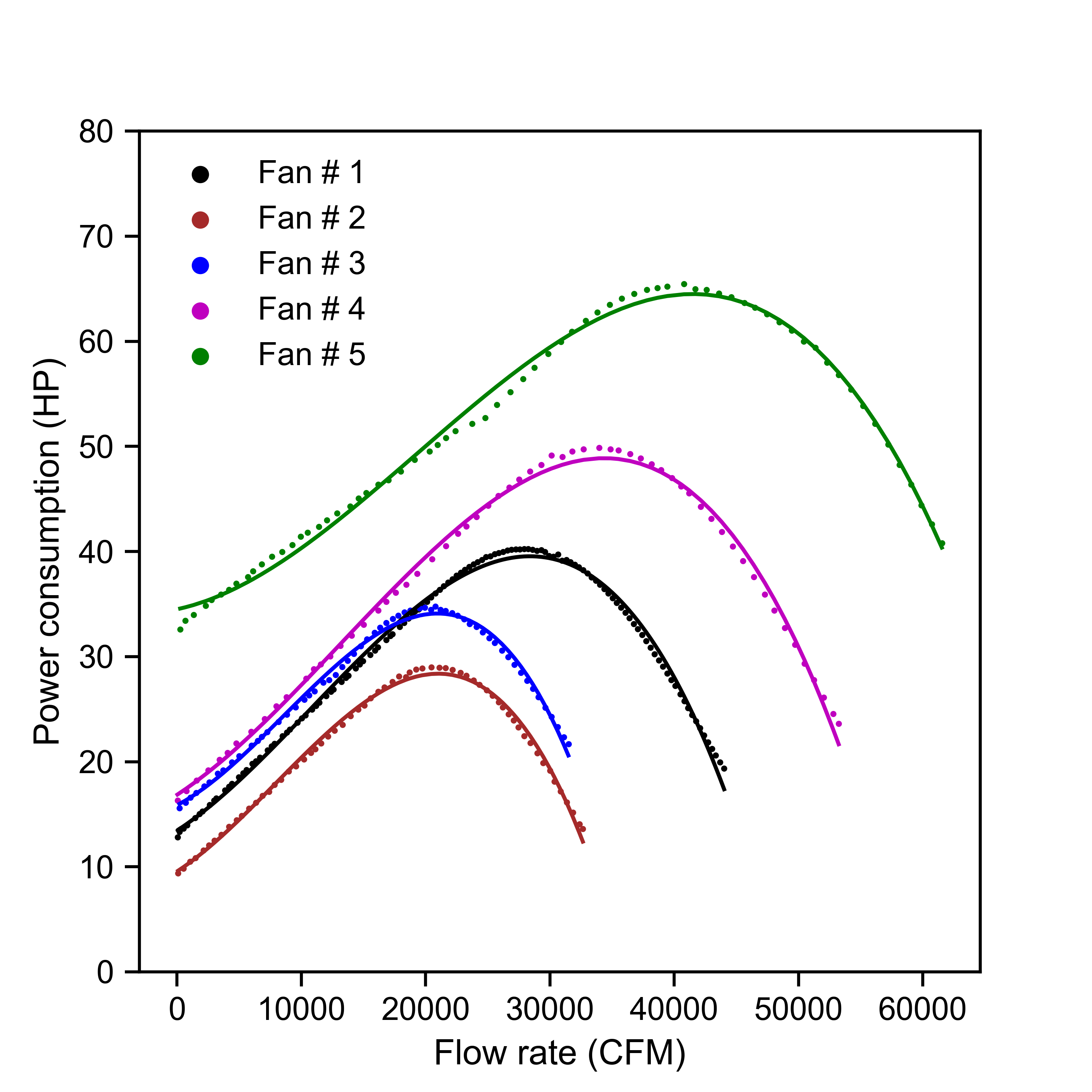}
    \caption{Identification results for energy performance models of supply fans in Bldg. A}
    \label{fig:fan performance curves}
\end{figure}

\Cref{fig:fan performance curves} shows the identification results of the energy performance of supply fans in Bldg. A. The points are the measured fan power consumption under various operation conditions at the commissioning stage after the fans were installed. The solid lines are the fan power consumption predicted using the polynomial model from \Cref{eq: fan power}. Fan specifications and the identified coefficients, $a_0$-$a_3$, and $R^2$ for each fan in Bldg. A are listed in \Cref{tab:identification results for fans}. Note that for Bldg. B and Bldg. C, only the rated air flow rate and rated power consumption are known and the measured power consumption at the commissioning stage was not recorded, which means the performance models can not be identified. In our study, the power consumption of fans in Bldg. B and Bldg. C is then estimated based on the energy performance model of the fan in Bldg. A, which has similar rated air flow rate and rated power consumption to the target fan. Specifically, the power consumption of fans in Bldg. B and Bldg. C can be estimated using \Cref{eq: fan power performance - adjusted}, where $p_{rated,1}$ and $p_{rated,2}$ are the rated power consumption of Fan 1 (energy performance model is known with the identified coefficients of $a_0$-$a_3$) and Fan 2 (energy performance model is unknown but have a similar rated power consumption to Fan 1).

\begin{equation}
    \label{eq: fan power performance - adjusted}
    p_{fan,2} = \frac{p_{rated,2}}{p_{rated,1}}(a_0 + a_1\times v + a_2\times v^2 + a_3\times v^3)
\end{equation}

After identifying the unknown coefficients in \Crefrange{eq: fresh air load regression model}{eq: aggregate coil cooling}, we can use \Crefrange{eq: zone-level equivalent coil cooling load}{eq: zone-level equivalent building cooling load} to estimate the cooling load of each zone. Then, we can obtain the aggregated estimate of the zone-level cooling load ($\hat{q}_{z,eb,t}^{i,j}$) and compare it with the measured building-level cooling load. The coefficient of variation of the root mean square error ($CV-RMSE$, \Cref{eq: cv rmse}) is used in our study to evaluate the accuracy of the estimation method at the building level. The $CV-RMSEs$ between the measured building-level cooling load and the aggregated estimate of zone-level cooling load for Buildings A-C are 24.4\%, 21.4\%, and 22.4\%, respectively. Note that when the $CV-RMSE$ is around 30\%, it means the developed model is calibrated and acceptable for engineering purposes \cite{Fan2017-pi, Reddy2007-mo}.

\begin{equation}
    \label{eq: cv rmse}
    CV-RMSE = \sqrt{\frac{\sum_{k=1}^{N}(y_k - \hat{y}_k)^2}{N}}/\frac{\sum_{k=1}^{N}y_k}{N}
\end{equation}

\begin{table}[htbp]
    \centering
    \caption{
    \label{tab:identification results for fans} Fan specifications and the identification results for fan energy performance models}
    \begin{tabular}{llllllll}
    \hline \hline
     & Rated air flow rate (CFM) & Rated power consumption (HP) & $a_0$ & $a_1$ & $a_2$ & $a_3$ & $R^2$ \\
     \hline
     Fan 1 & 30000 & 40.83 & 13.45 & 0.00077 & $4.30e{-8}$ & $-1.33e{-12}$ & 0.996 \\
     Fan 2 & 22000 & 30.7 & 9.56 & 0.00077 & $5.48e{-8}$ & $-2.32e{-12}$ & 0.995 \\
     Fan 3 & 22000 & 30.7 & 15.86 & 0.00062 & $6.57e{-8}$ & $-2.56e{-12}$ & 0.994 \\
     Fan 4 & 37000 & 48.1 & 16.85 & 0.00077 & $3.64e{-8}$ & $-9.25e{-13}$ & 0.995 \\
     Fan 5 & 44000 & 63.3 & 34.51 & 0.00025 & $4.03e{-8}$ & $-6.95e{-13}$ & 0.994 \\
    \hline
    \hline
    \end{tabular}%
\end{table}

\subsection{Building-level and AHU-level energy performance}
\begin{figure}[htbp]
    \centering
    \includegraphics[width=1\textwidth]{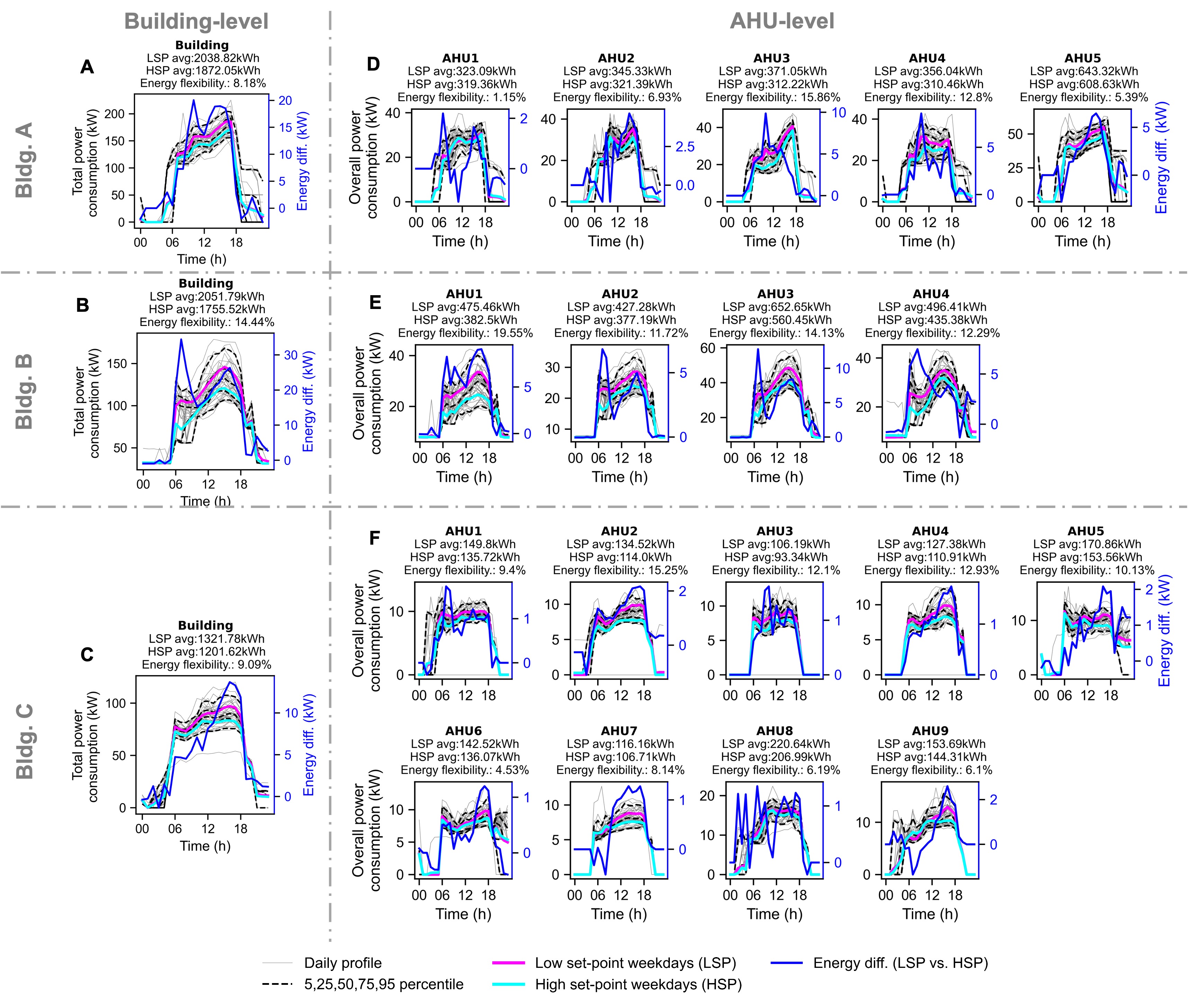}
    \caption{Building-level and AHU-level energy performance, including energy use and energy flexibility. (A-C) Building level; (D-F) AHU level}
    \label{fig: building and ahu level energy performance}
\end{figure}

After identifying the unknown coefficients in \Crefrange{eq: fresh air load regression model}{eq: fan power}, we can use the set of models in \Crefrange{eq: zone-level equivalent coil cooling load}{eq: zone-level equivalent overall electrical load} to estimate the overall electrical load of each zone. Based on the zone-level electrical load, the AHU-level and building-level electrical loads can then be calculated. Figure 7 shows the overall electrical load at AHU and building levels for the three target buildings. At the building level, it can be observed from \Cref{fig: building and ahu level energy performance} A-C that all three target buildings can provide a certain amount of energy flexibility (blue) by increasing indoor air temperature set-point by 2\degree F (1.1\degree C). The energy flexibility of Bldgs. A-C calculated based on \Cref{eq: performance index - energy flexibility} are 8.18\%, 14.44\%, and 9.09\%, respectively. Moreover, the shapes of the building-level energy flexibility profiles are different among buildings; the energy flexibility of Bldg. A and Bldg. B reaches its peak in the morning while the energy flexibility of Bldg. C reaches its peak in the afternoon.

In addition to the flexibility in total HVAC-related electrical load, we can also evaluate the flexibility in thermal cooling load and fan power consumption for different buildings by using our load disaggregation method. As listed in \Cref{tab: building level energy performance}, the flexibility in fan power consumption is lower than the flexibility in thermal cooling load. Specifically, the flexibility in thermal cooling load is 20.9\%, 29.4\%, and 18.0\%, respectively, while the flexibility in fan power consumption is 2.0\%, 4.5\%, and 4.7\%, respectively. Also, we can find that the fan power consumption accounts for around 60\% of the total HVAC-related electrical load for each building. Note that the relationship between these types of energy consumption can be described by using \Cref{eq: zone-level equivalent overall electrical load}.

\begin{figure}[htbp]
    \centering
    \includegraphics[width=1\textwidth]{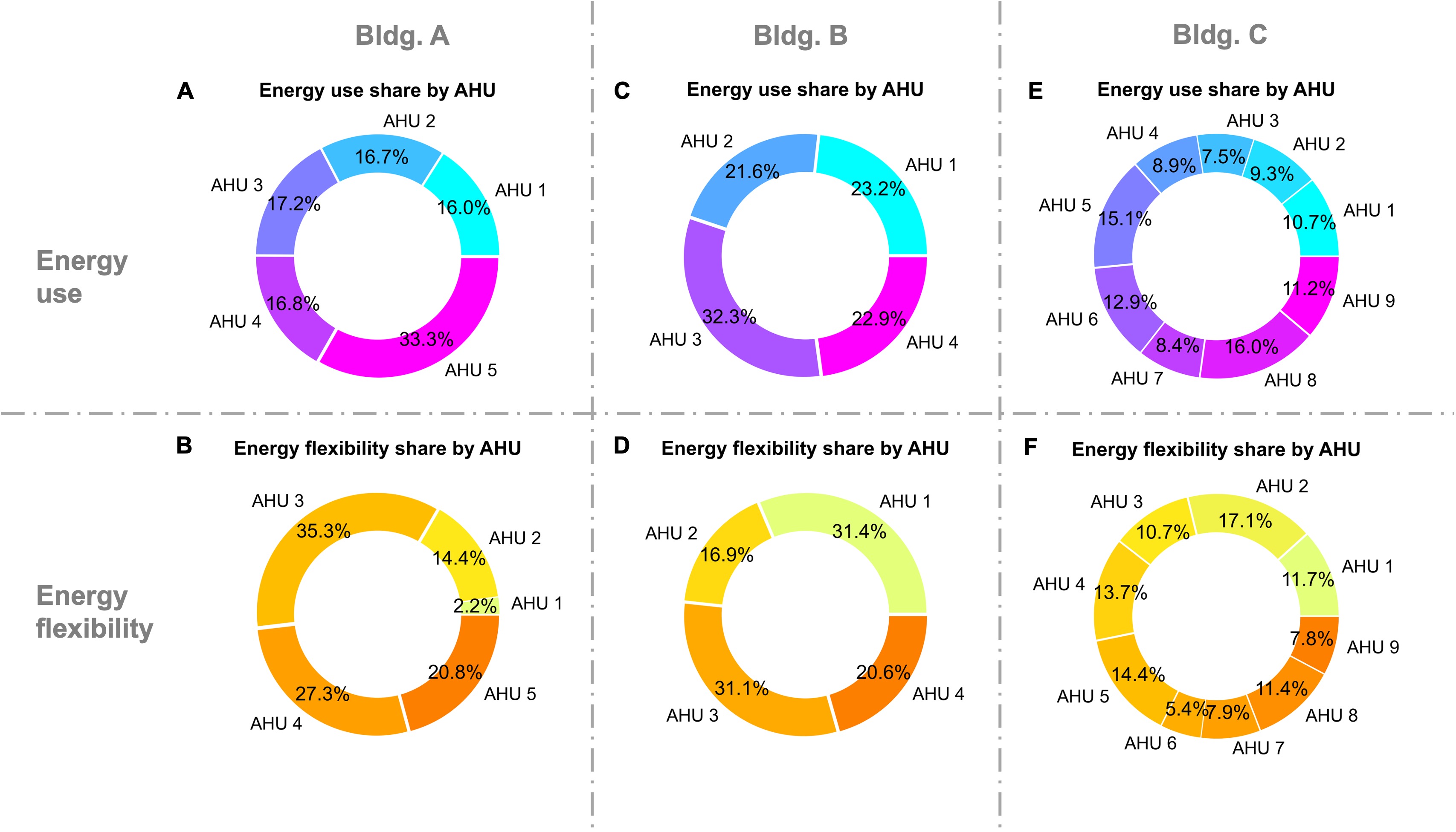}
    \caption{Distribution of AHU-level energy use share (A, C, and E) and energy flexibility share (B, D, and F) in each building}
    \label{fig:AHU level energy use share and ef share}
\end{figure}

At the AHU level, it can be found from \Cref{fig: building and ahu level energy performance} D-F that the ranges of the AHU-level energy flexibility for Bldgs. A-C are 1.15\% - 15.86\%, 11.72\% - 19.55\%, 4.53\% - 15.25\%, respectively. The shapes of the AHU-level energy flexibility profiles are also heterogeneous. Apart from the energy flexibility calculated by \Cref{eq: performance index - energy flexibility}, we can also analyze the energy use and energy flexibility shares of each AHU inside a building using \Cref{eq: performance index - energy flexibility share}. As shown in \Cref{fig:AHU level energy use share and ef share}, AHU 3 in Bldg. A consumes 17.2\% of the total energy use but provides the highest share of energy flexibility, 35.3\%. In contrast, AHU 1 in Bldg. A consumes 16.0\% of the total energy use, but only accounts for 2.2\% of the total energy flexibility. Likewise, AHU 6 in Bldg. C consumes 12.9\% of the total energy use, but only accounts for 5.4\% of the total energy flexibility. AHUs in Bldg. B have a smaller difference in energy flexibility share than Bldg. A and Bldg. C.

\begin{table}[ht]
    \centering
    \caption{
    \label{tab: building level energy performance} Building-level flexibility in different types of energy consumption, including thermal and electrical cooling load, fan power use, and total HVAC-related electrical load. (The relationship between these energy consumption types is described using \Cref{eq: zone-level equivalent overall electrical load}; LSP - low set-point; HSP - high set-point.)}
    \begin{adjustbox}{width=\textwidth}
    \begin{tabular}{lllllll}
    \hline \hline
     & & Thermal load for cooling & Electrical load for cooling & Electrical load for fan & Total HVAC-related electrical load & Fan power \% \\
    \hline
    \multirow{4}{*}{Bldg. A} & Avg. daily value on LSP days (kWh) & 2841.9  & 739.1 &  1299.7 & 2038.8 & 63.8\%  \\
                         & Avg. daily value on HSP days (kWh) & 2247.6 & 598.7 & 1273.4 & 1872.1 & 68.0\%   \\
                         & Flexibility (\%) & 20.9\% & 19.0\%  & 2.0\% & 8.2\% &  \\
    \hline
    \multirow{4}{*}{Bldg. B} & Avg. daily value on LSP days (kWh) & 3222.6  & 840.6 & 1211.2 & 2051.8 & 59.0\%  \\
                     & Avg. daily value on HSP days (kWh) & 2274.8 & 598.8 & 1156.7 & 1755.5 & 65.9\%   \\
                      & Flexibility (\%) & 29.4\% & 28.8\% & 4.5\% & 14.4\% &  \\
    \hline 
    \multirow{4}{*}{Bldg. C} & Avg. daily value on LSP days (kWh) & 1993.0 & 520.8  &  801.0 & 1321.8 & 60.6\%  \\
                     & Avg. daily value on HSP days (kWh) & 1634.9 & 438.5 & 763.1 & 1201.6 & 63.5\%   \\
                      & Flexibility (\%) & 18.0\% & 15.8\% & 4.7\% & 9.1\% &  \\
    \hline \hline
    \end{tabular}%
    \end{adjustbox}
\end{table}

\subsection{Zone-level energy performance}
\begin{figure}[htbp]
    \centering
    \includegraphics[width=0.7\textwidth]{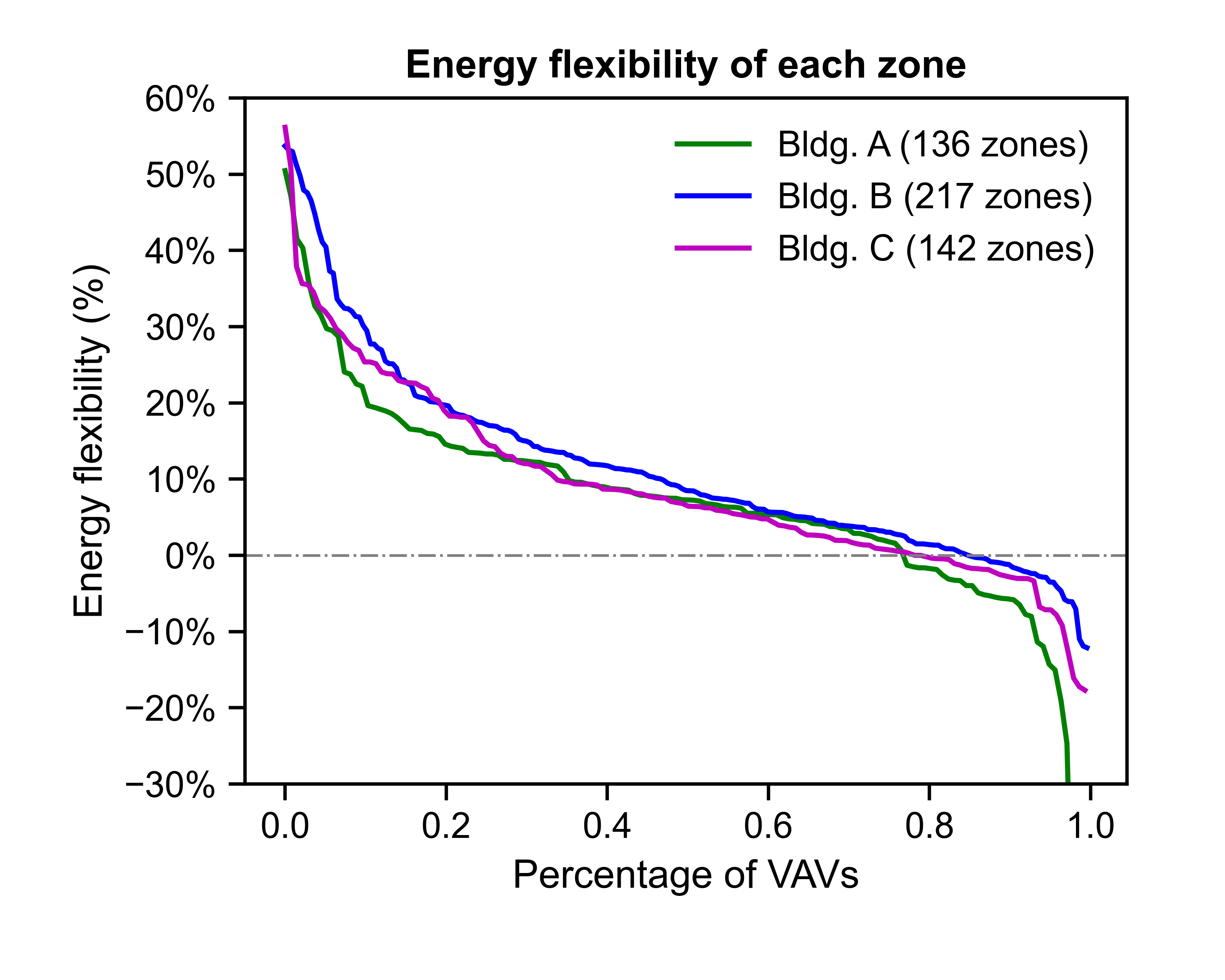}
    \caption{Distribution of zone-level energy flexibility in each building from highest to lowest ones. There are 3 zones in Bldg. A with an energy flexibility value of less than -30\%.}
    \label{fig: energy flexibility of each zone}
\end{figure}

One of the major contributions of this work is we can investigate the zone-level energy use and energy flexibility by using the experimental data and proposed load disaggregation approach. The zone-level energy flexibility, energy use share, and energy flexibility share are analyzed in this subsection. It can be found from \Cref{fig: energy flexibility of each zone} that not all zones in target buildings provide positive energy flexibility, which means some zones consume more overall electrical load while the indoor air temperature set-point is increased by 2\degree F. The percentages of the zones with negative energy flexibility for Bldgs. A-C are 22.79\%, 14.74\%, and 20.42\%, respectively. The highest energy flexibility provided by a single zone for Bldgs. A-C is 50.47\%, 53.71\%, and 56.16\%, respectively.

\begin{figure}[htbp]
    \centering
    \includegraphics[width=.85\textwidth]{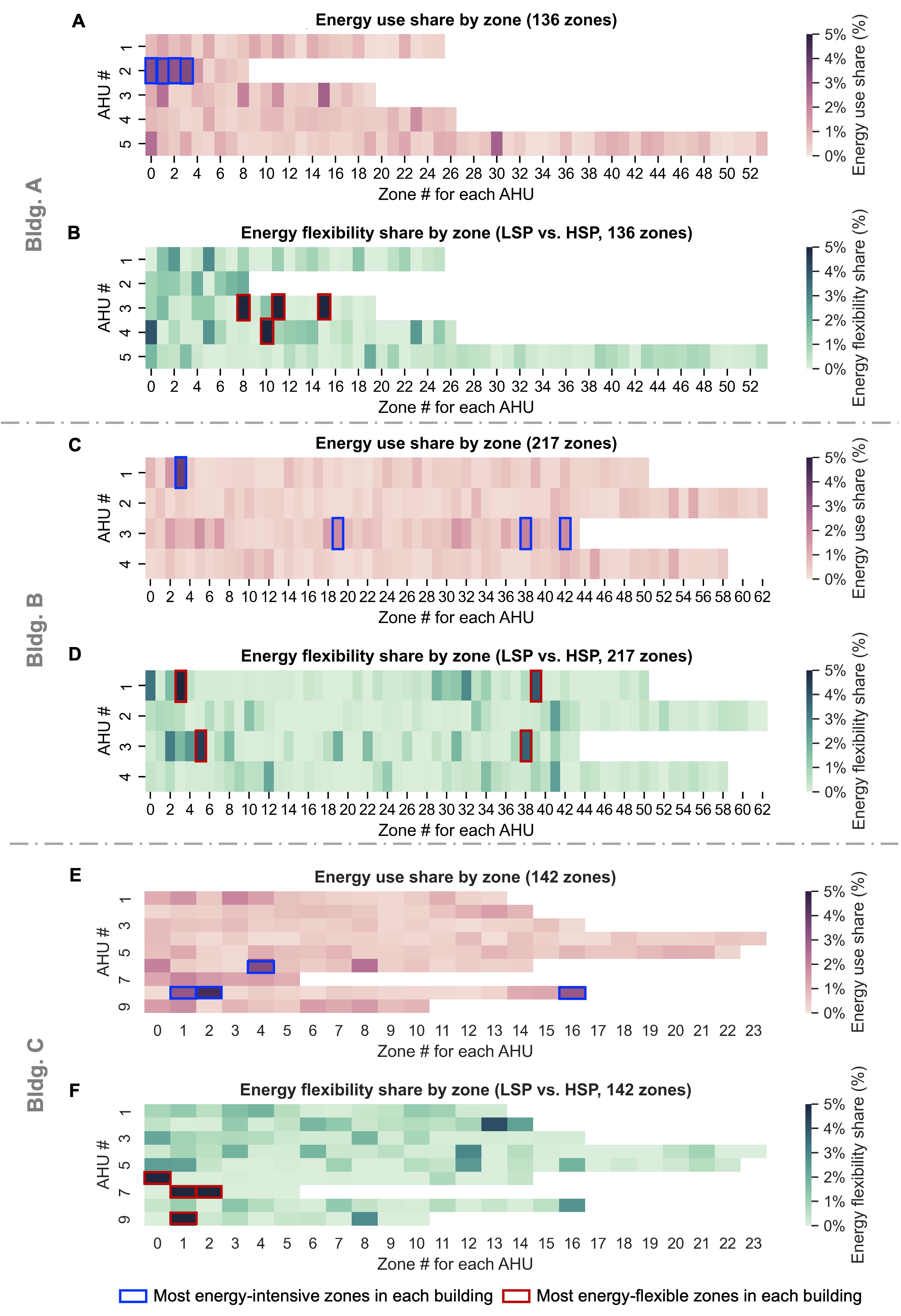}
    \caption{Heat maps of energy use share and energy flexibility share by zone in each target building}
    \label{fig:zone level energy performance}
\end{figure}

We make three important observations from the heat maps of energy use share and energy flexibility share by zone for each building shown in \Cref{fig:zone level energy performance}. First, we observe from \Cref{fig:zone level energy performance} that both energy use share and energy flexibility share are heterogeneous across zones and AHUs in each building. Furthermore, the heterogeneity of energy use and energy flexibility across zones can be quantified using the Gini coefficient, i.e., \Cref{eq: performance index - gini index}. \Cref{fig:gini coefficient} shows the energy use (EU) heterogeneity and energy flexibility (EF) heterogeneity of Bldgs. A-C. Compared to Bldg. A and Bldg. B, Bldg. C has the largest Gini coefficients in terms of both energy use and energy flexibility, which indicates Bldg. C has the highest diversity in both energy use and energy flexibility. Bldg. B has a smaller Gini coefficient of energy use than Bldg. A but a larger Gini coefficient of energy flexibility than Bldg. A.

\begin{figure}[htbp]
    \centering
    \includegraphics[width=1\textwidth]{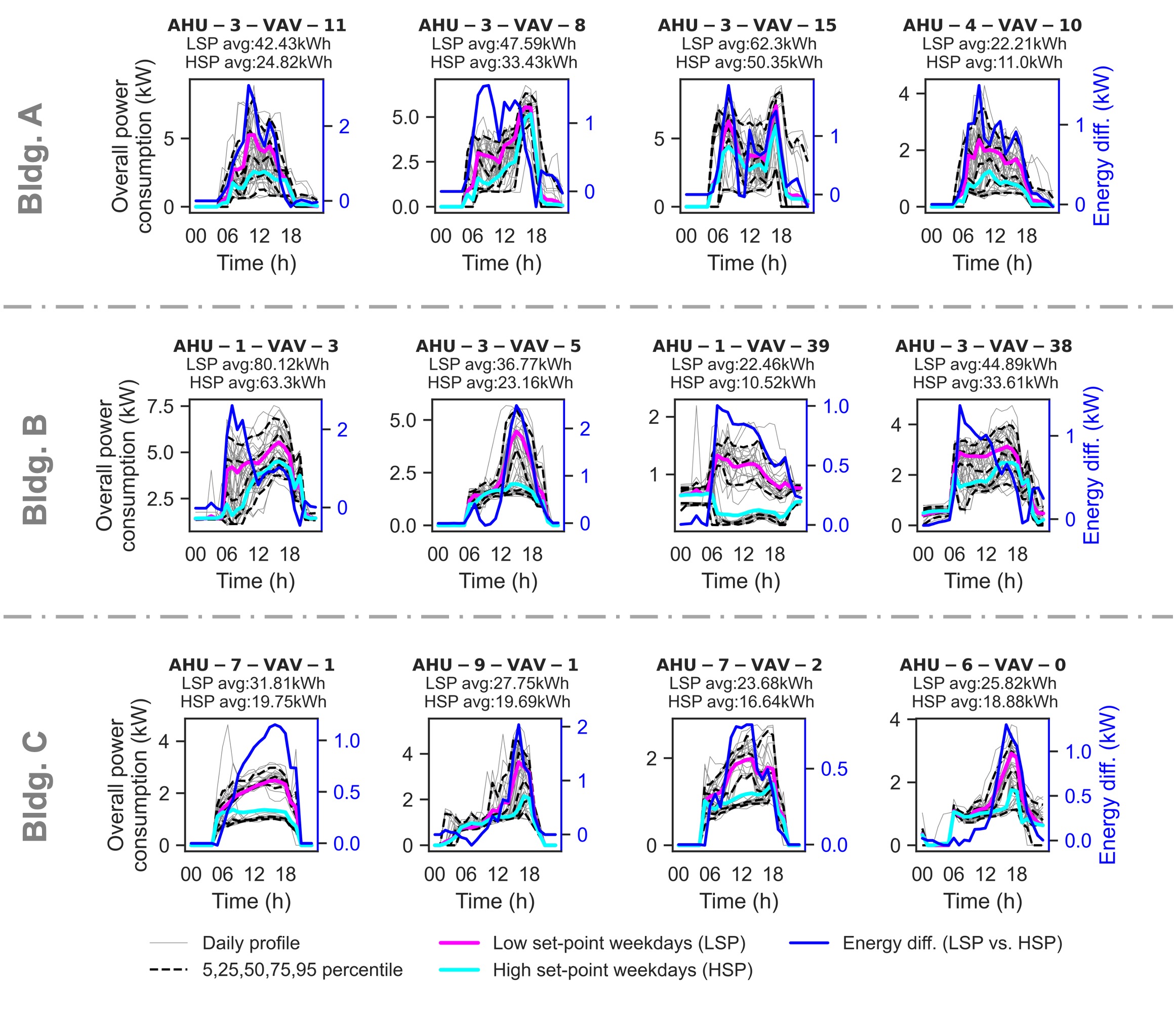}
    \caption{Hourly energy profiles of the most energy-flexible zones}
    \label{fig: most energy flexible zones}
\end{figure}

\begin{figure}[htbp]
    \centering
    \includegraphics[width=1\textwidth]{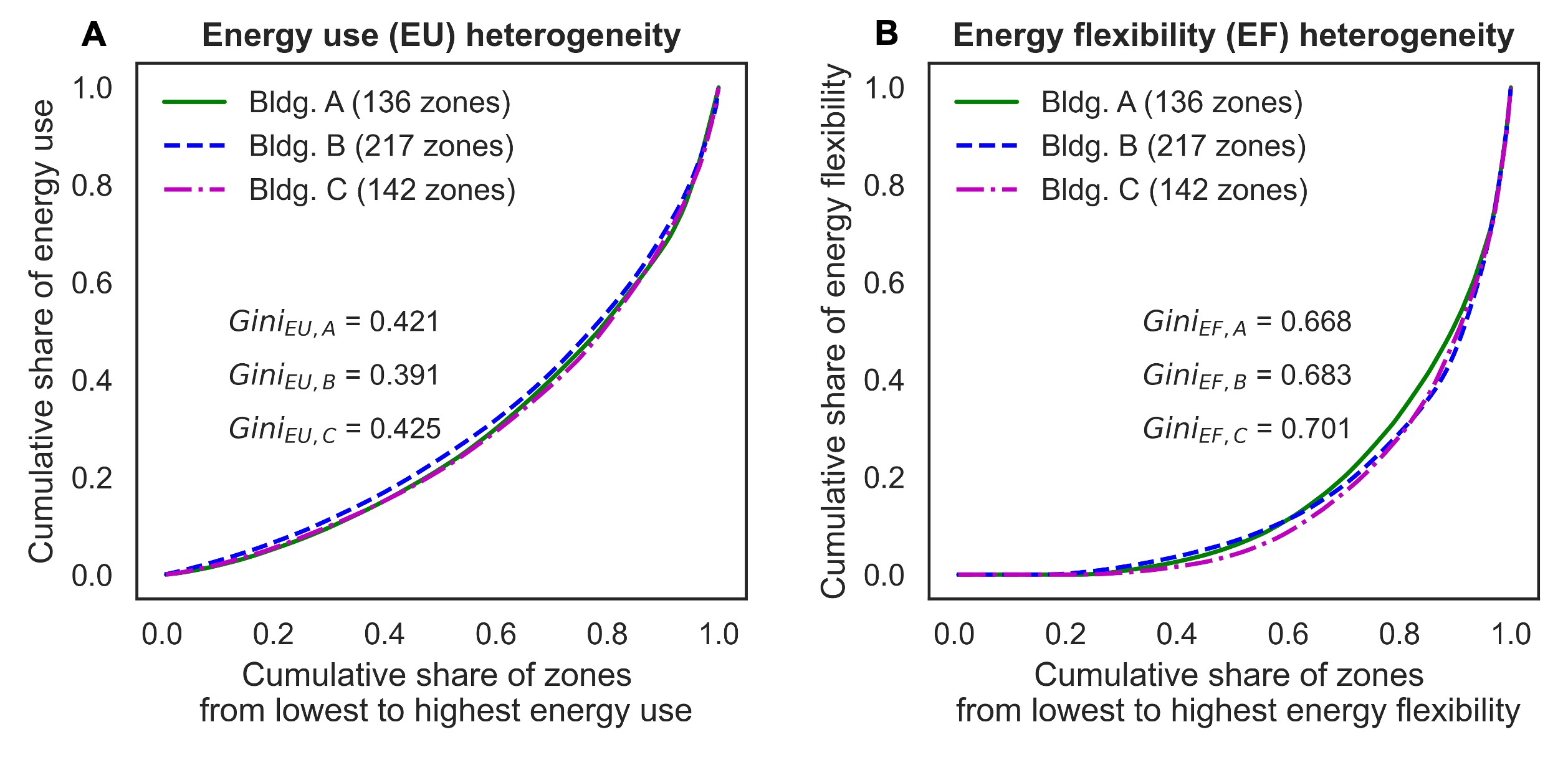}
    \caption{Comparison of the heterogeneities of energy use and energy flexibility using Lorenz curves and Gini coefficients}
    \label{fig:gini coefficient}
\end{figure}

Second, the most energy-flexible zones are not necessarily the most energy-intensive zones for each building. Blue and red boxes in \Cref{fig:zone level energy performance} denote most energy-intensive and energy-flexible zones in each building, respectively. The zones with blue boxes are not the same as the zones with red boxes. The overall energy use and energy flexibility profiles of the most energy-flexible zones are also plotted in \Cref{fig: most energy flexible zones}.

\begin{figure}[htbp]
    \centering
    \includegraphics[width=1\textwidth]{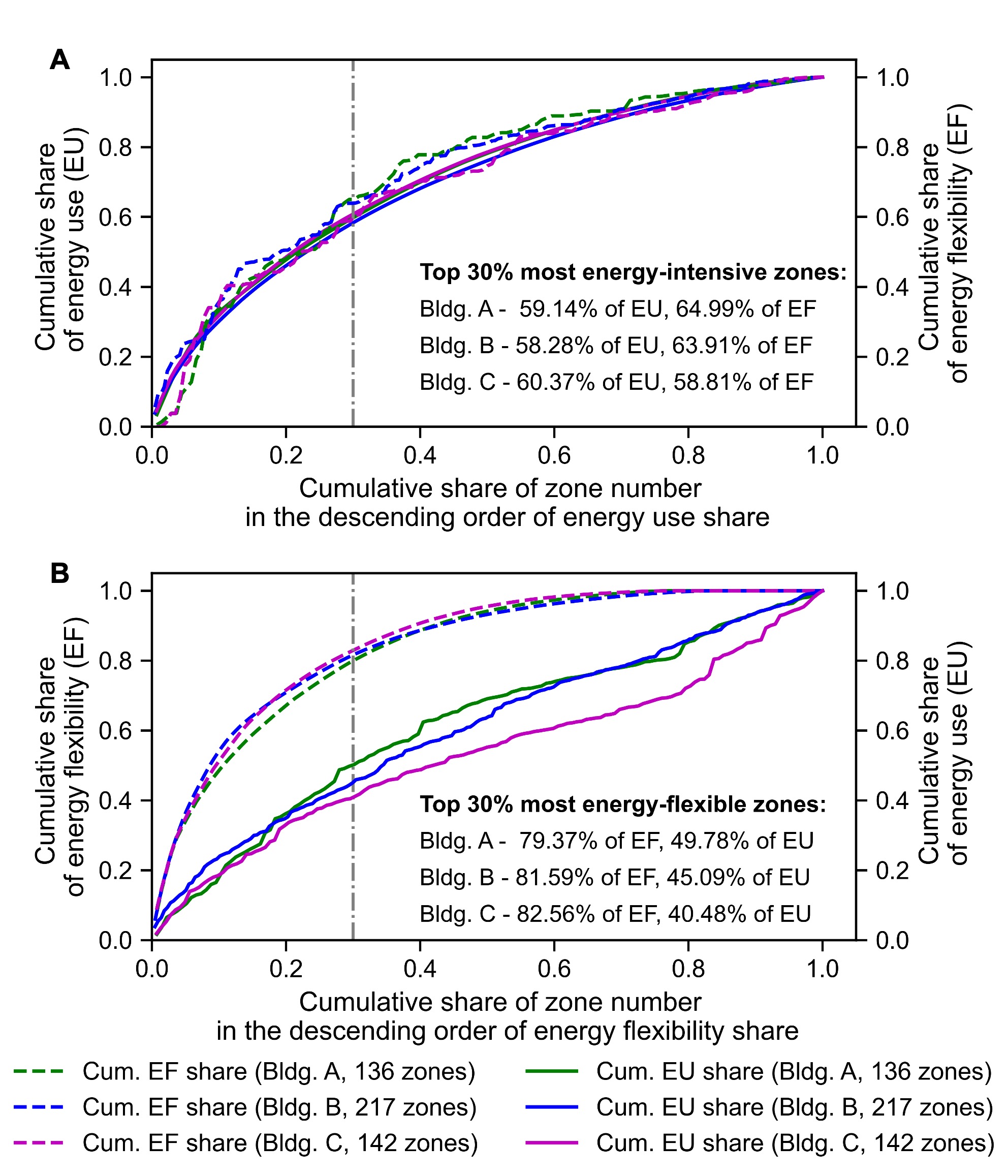}
    \caption{Distribution of energy use and energy flexibility across zones in target buildings and the identification of most energy-intensive and energy-flexible zones}
    \label{fig:cum energy use vs cum energy flexibility}
\end{figure}

Third, a small number of zones account for a large amount of energy use and energy flexibility. \Cref{fig:cum energy use vs cum energy flexibility}-A shows the cumulative shares of energy use (left Y-axis) and energy flexibility (right Y-axis) when X-axis represents the cumulative share of the number of zones in the descending order of the energy use share. It can be found that the top 30\% of most energy-intensive zones account for 59.14\%, 58.28\%, and 60.37\% of the total energy use and represent 64.99\%, 63.91\%, and 58.81\% of the total energy flexibility in Bldgs. A-C, respectively. Similarly, \Cref{fig:cum energy use vs cum energy flexibility}-B shows the cumulative shares of energy flexibility (left Y-axis) and energy use (right Y-axis) when X-axis represents the cumulative share of the number of zones in the descending order of the energy flexibility share. The top 30\% of most energy-flexible zones account for 79.37\%, 81.59\%, and 82.56\% of the total energy flexibility and consume 49.78\%, 45.09\%, and 40.48\% of the total energy use in Bldgs. A-C, respectively.

\subsection{Zone-level thermal performance}
\begin{figure}[htbp]
    \centering
    \includegraphics[width=.85\textwidth]{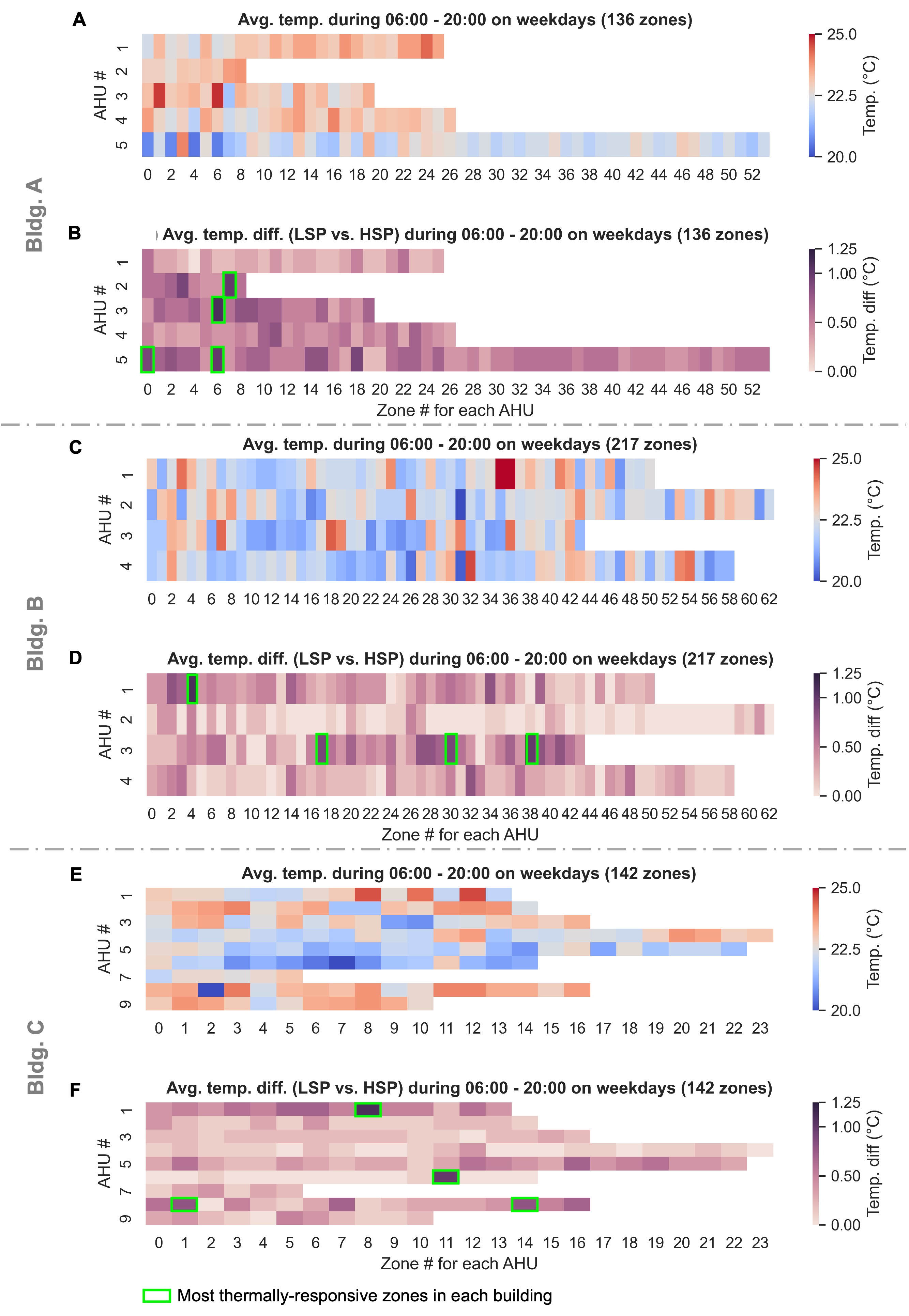}
    \caption{Heat maps of indoor air temperature and thermal impacts of DR control strategy across zones in target buildings}
    \label{fig:zone-level thermal impact}
\end{figure}

\begin{figure}[htbp]
    \centering
    \includegraphics[width=1\textwidth]{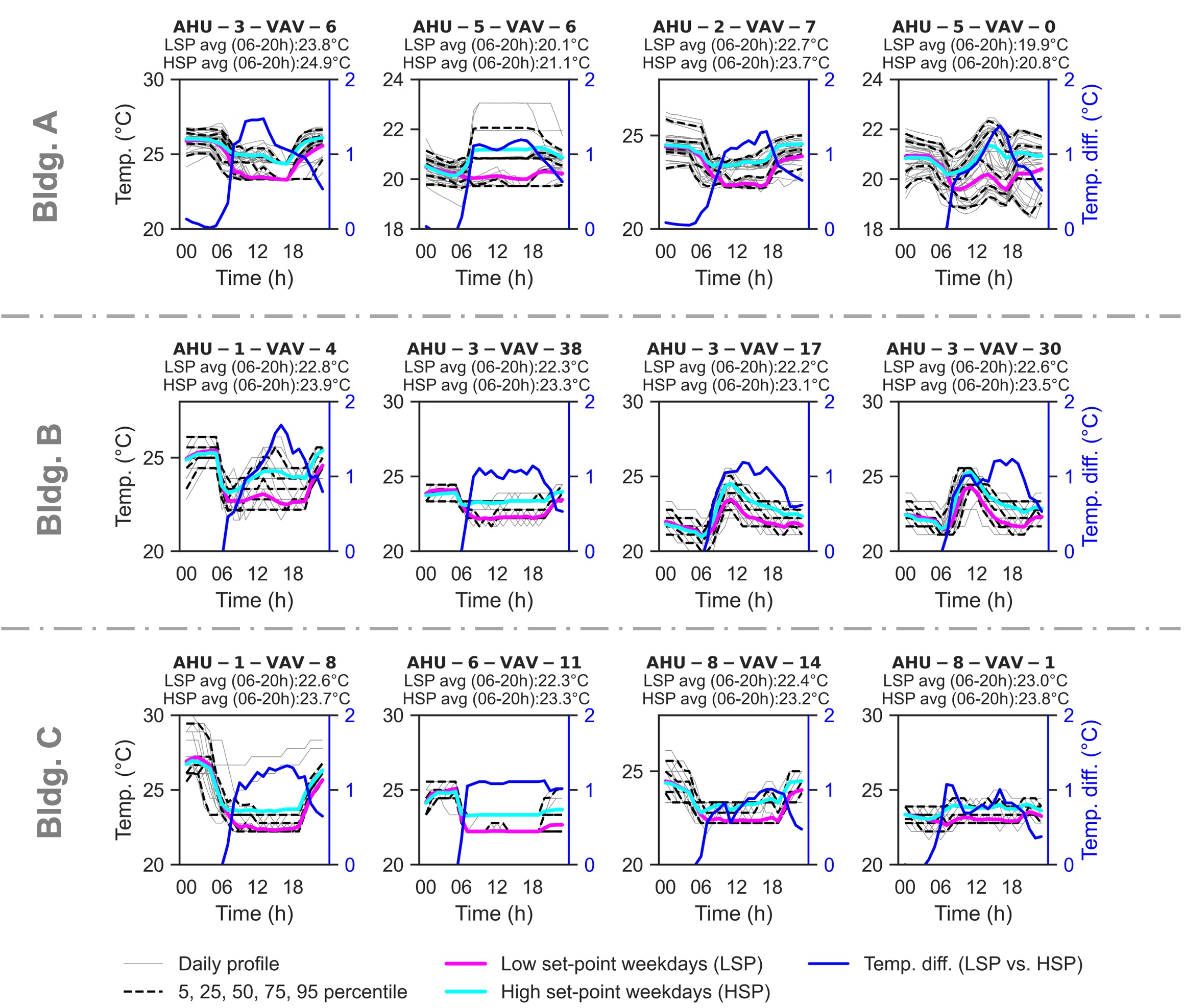}
    \caption{Temperature profiles of the most thermally responsive zones in each building}
    \label{fig:most thermally responsive zones}
\end{figure}

\begin{figure}[htbp]
    \centering
    \includegraphics[width=0.7\textwidth]{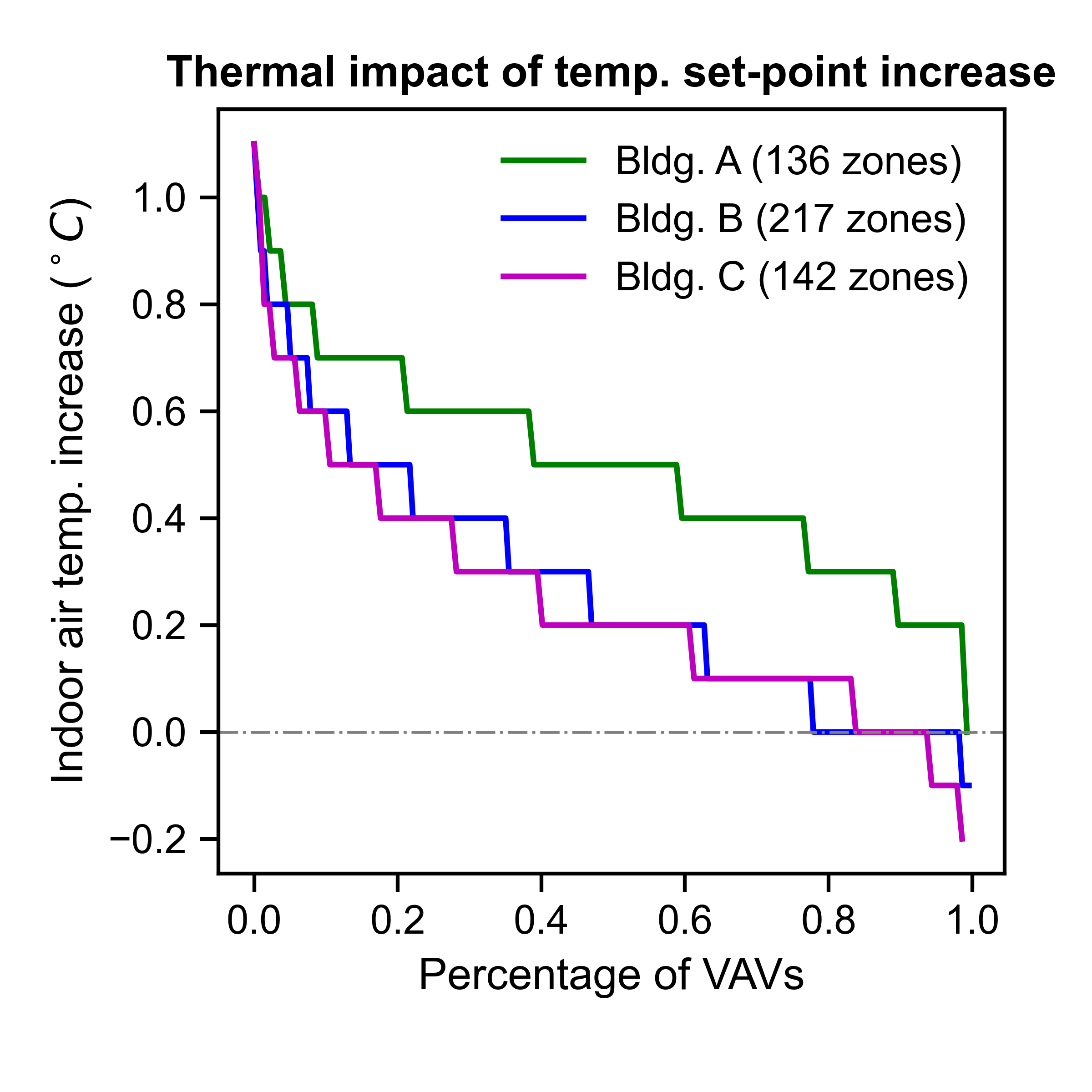}
    \caption{Distribution of zone-level thermal impact from highest to lowest temperature increase}
    \label{fig:thermal impacts of temp setpoint increase}
\end{figure}

In addition to the energy performance, the impacts of the DR control strategy (i.e., temperature set-point adjustment) on indoor air temperature were also investigated in our study. \Cref{fig:zone-level thermal impact} (A, C, and E) show the average indoor air temperature from 06:00 to 20:00 during the experimentation for each building. \Cref{fig:zone-level thermal impact} (B, D, and F) show the average temperature difference between the low set-point case and the high set-point case (DR case) for each building. Green boxes in \Cref{fig:zone-level thermal impact} represent the most thermally-responsive zones in each building. The daily temperature profiles of these zones are plotted in \Cref{fig:most thermally responsive zones}. \Cref{fig:thermal impacts of temp setpoint increase} shows the distributions of zone-level indoor air temperature increase for each target building. It is ordered from the highest to lowest temperature increase, and X-axis represents the percentage of the VAVs in each building.

Without considering the temperature set-point increase, the indoor air temperatures are found to be heterogeneous across the zones in the same building and from building to building. Moreover, the symptom of over-cooling can be found in many rooms, i.e., the indoor air temperatures are below the cooling set-point, and the over-cooling degree varies across the zones. As shown in \Cref{fig:most thermally responsive zones}, during the low set-point (23.3\degree C) days, the measured average indoor air temperature in most zones is below the set-point to a different extent.

Regarding the thermal response to temperature increase, the following insights can be observed. First, not all zones have a certain increase in indoor air temperature when the cooling set-points are increased by 2\degree F (1.1\degree C) as shown in \Cref{fig:thermal impacts of temp setpoint increase}. The percentages of zones with an indoor air temperature decrease are 0\%, 1.38\%, and 4.93\% for Bldgs. A-C, respectively. Second, the changes in indoor air temperature due to the set-point adjustment (1.1\degree C) are limited, and most rooms in Bldgs. A-C had a temperature change less than the set-point adjustment. Third, the effects of temperature set-point adjustment on indoor air temperature are heterogeneous. Compared to Bldgs. B and C, Bldg. A has a stronger response to the temperature set-point increase. The mean values of indoor air temperature increase are 0.502\degree C, 0.274\degree C, and 0.250\degree C for Bldgs. A-C, respectively.

\newpage
\section{Discussion}

\subsection{A ‘virtual’ smart power meter to provide zone-level energy performance insights in multi-zone buildings}

We developed a non-intrusive load disaggregation method to analyze the zone-level energy performance in multi-zone commercial buildings using data from IoT sensors and physics-based data-driven models. The key idea of this work is to develop a ‘virtual’ smart power meter to provide high-granularity insights into zone-level energy performance, including energy use and energy flexibility. To obtain accurate equivalent power consumption of each zone, rather than the direct space cooling load in Ref. \cite{Balaji2013-tu}, the following challenges were addressed by this study. First, we developed regression models to address the issue of mismatched cooling loads at different levels, as illustrated in \Cref{fig: different levels of cooling loads}. The essence of the regression models is to adjust the space cooling load and make the aggregate adjusted space cooling load of all zones equal to the measured building-level load. \Cref{fig:regression models_bldg_A} shows that the regression models can effectively address the issue of mismatched cooling loads. After estimating the adjusted space cooling load, we can convert it to the electrical load based on the real-time COP for cooling generation in the district cooling system. Second, compared to the study of \cite{Balaji2013-tu}, the fan power consumption was considered in our study. The fan power is not sub-metered in our experimental testbed. We developed a data-driven polynomial fan power consumption model and trained the model by using historical operation data. After training, we can estimate the real-time fan power consumption based on the measured air flow rate. As listed in \Cref{tab: building level energy performance}, the fan power consumption represents around 60\% of the HVAC-related electrical load at the building level in target buildings, which indicates that the consideration of the fan power consumption is necessary when evaluating the energy performance of HVAC systems. The estimated fan power consumption can be proportionally divided according to the air flow rate of each zone. 

The proposed load disaggregation models are physically interpretable, simple, and scalable. We tested the models on three different campus buildings, and the results show that the proposed models can effectively address the above two challenges and help us estimate the zone-level energy performance. Furthermore, we can investigate the zone-level energy flexibility when combining the numerical load disaggregation technique with the implementation of the DR control strategy.

\subsection{Heterogeneities of energy use and energy flexibility}

With the repetitive implementation of cooling temperature set-point adjustment and the proposed data-driven load disaggregation technique, one of the major findings of this work is that heterogeneities of energy use and energy flexibility exist across not only buildings but also AHUs and zones. AHUs and zones in a building do not identically respond to the cooling temperature set-point adjustment, and some AHUs and zones have larger energy flexibility than others.

At the building level, the flexibility in total electrical load provided by Bldgs. A-C is 8.18\%, 14.44\%, and 9.09\%, respectively, as shown in \Cref{fig: building and ahu level energy performance}. Additionally, as indicated in \Cref{tab: building level energy performance}, the flexibility in thermal cooling load (18.0\% - 29.4\%) is larger than the flexibility in fan power consumption (2.0\% - 4.7\%). This finding contrasts with several previous experimental studies \cite{MacDonald2020-ri, Keskar2022-nj, Beil2015-ip, Afshari2017-qj} where the flexibility of HVAC systems was mainly from the ventilation fans. The reason is that in most AHUs in our target buildings, supply air pressure and temperature reset strategies (i.e., `Trim-and-Respond' logic) \cite{Raftery2018-xe}  were implemented, while the supply air temperature in prior studies is maintained fixed.

At the AHU level, we can find from \Cref{fig:AHU level energy use share and ef share} that AHU 3 in Bldg. A consumes 17.2\% of the total energy use but provides the highest share of energy flexibility, 35.3\%. In contrast, AHU 1 in Bldg. A consumes 16.0\% of the total energy use, but only accounts for 2.2\% of the total energy flexibility. At the zone level, not all zones provide positive energy flexibility as shown in \Cref{fig: energy flexibility of each zone}, which means some zones consume more overall electrical load while the indoor air temperature set-point is increased by 2\degree F. In addition, both energy use and energy flexibility are heterogeneous across zones and AHUs in each building as shown in \Cref{fig:zone level energy performance}. The heterogeneities of energy use and energy flexibility across zones can be further quantified using the Gini coefficient, a commonly used metric in economics to describe income inequality. \Cref{fig:gini coefficient} shows that the proposed Gini coefficient can effectively quantify the diversities of energy use and energy flexibility across zones. Moreover, we can also find that the most energy-flexible zones are not necessarily the most energy-intensive zones for each building as shown in \Cref{fig:zone level energy performance}. Lastly, a small number of zones account for a large amount of energy use and energy flexibility. As shown in \Cref{fig:cum energy use vs cum energy flexibility}, the top 30\% of most energy-intensive zones account for around 60\% of the total energy use; and the top 30\% of most energy-flexible zones provide around 80\% of the total energy flexibility. We found that the majority of the most energy-flexible zones within a building are large zones with dynamic occupant densities, such as 1) expansive shared research laboratories, maker laboratories, and offices with open plans, 2) large classrooms with windows, and 3) communal areas like corridors and cafeteria spaces.

This finding from our work enables the electric grid or district energy system operators to rethink the implementation of DR control strategies for multi-zone commercial buildings. Conventionally, the electric grid or district energy system operators consider which buildings in a district they should prioritize controlling for DR. The proposed data-driven load disaggregation technique enables them to regard the fleet of buildings as a fleet of AHUs or zones, which helps unlock the possibility for targeted demand flexibility strategies that balance zone-by-zone energy reductions with zone-by-zone costs to occupants. For example, the occupants with a higher thermal tolerance could be relocated to rooms with greater flexibility potential.

\subsection{Heterogeneities of indoor air temperature, over-cooling degree and DR thermal impact}

After the repetitive implementation of cooling temperature set-point adjustment, the effects of the DR control strategy (i.e., temperature set-point increase) on indoor air temperature were also investigated in this work. The reason for analyzing the zone-level thermal impact of the DR control strategy is that it is the zones where the thermal costs of DR control are paid. 

Without DR measures, the indoor air temperatures are first found to be heterogeneous across the zones in the same building and from building to building. Moreover, the symptom of over-cooling can be found in many rooms, i.e., the indoor air temperatures are below the cooling set-points, and the over-cooling degree varies across the zones as shown in \Cref{fig:zone-level thermal impact}. Regarding the thermal response to DR measures, not all zones have a certain increase in indoor air temperature when the cooling set-points are increased by 2\degree F. Finally, the effects of temperature set-point adjustment on indoor air temperature are limited and heterogeneous. As shown in \Cref{fig:thermal impacts of temp setpoint increase}, most rooms in target buildings had a temperature change less than the set-point adjustment, and the changes are found to be heterogeneous.

The hourly temperature profiles are dynamic and vary from zone to zone. Furthermore, the thermal impact of the DR measures was heterogeneous. Due to these reasons, building operators need to consider what will really happen to the occupants when they implement DR measures in practice. Therefore, zone-level thermal dynamics need to be predicted before rolling out DR programs, which would be the focus of our future work.

\section{Conclusion}
In this paper, we investigated the impacts of DR control strategies (i.e., temperature set-point adjustment) on zone-level energy and thermal performance in multizone commercial buildings. This was accomplished by developing a data-driven non-intrusive load disaggregation method and designing a scalable DR experimental method. We draw three major conclusions from this study.

First, the key idea of the proposed non-intrusive load disaggregation method is to develop a ‘virtual’ power meter for each zone in multizone commercial buildings. It was tested on three different educational buildings in our study, and the results showed that it can effectively disaggregate the metered building-level cooling load into the zone-level equivalent electrical load with the consideration of fan power consumption. The models used for load disaggregation are physically interpretable, simple, and scalable. With the widespread IoT sensors in today’s commercial buildings, this load disaggregation technique can be used to provide building operators with high-granular insights into zone-level energy performance, including energy use and energy flexibility. 

Second, regarding energy performance, we found that heterogeneities of energy use and energy flexibility existed across not only buildings but also AHUs and zones. AHUs and zones in a building did not identically respond to the cooling temperature set-point adjustment, and some AHUs and zones had larger energy flexibility than others. Moreover, we found a small number of zones accounted for a large amount of energy use and energy flexibility. For the three tested buildings, the top 30\% of most energy-intensive zones account for around 60\% of the total energy use; and the top 30\% of most energy-flexible zones provide around 80\% of the total energy flexibility. These findings enable the electric grid or district energy system operators to rethink the implementation of DR control strategies for multi-zone commercial buildings. Conventionally, they need to decide which buildings they should prioritize providing energy flexibility during on-peak hours. The proposed load disaggregation technique enables them to regard the controlled entities as a fleet of AHUs or zones instead of a fleet of buildings and helps unlock the possibility for targeted demand flexibility strategies that balance zone-by-zone energy reduction with zone-by-zone costs to occupants. For example, the occupants with a higher thermal tolerance could be relocated to rooms with greater flexibility potential during on-peak hours.

Third, regarding thermal performance, we found that indoor air temperatures were heterogeneous across zones. The symptom of over-cooling can also be found in many zones and the over-cooling degrees varied across these zones. Moreover, the effects of temperature set-point adjustment on indoor air temperature were found to be limited and heterogeneous. These findings suggest that building operators need to consider the real thermal impact on the occupants when they implement DR measures. Zone-level thermal dynamics, therefore, need to be predicted before rolling out DR programs.

\section{Acknowledgments}
Funding for this research was supported by Stanford Land, Buildings and Real Estate, and TotalEnergies SE. The authors thank several past and current members of the Stanford Cooler project for valuable discussions and contributions that made the experiments reported on in this paper possible.

\bibliographystyle{elsarticle-num}

\begin{thebibliography}{10}
\expandafter\ifx\csname url\endcsname\relax
  \def\url#1{\texttt{#1}}\fi
\expandafter\ifx\csname urlprefix\endcsname\relax\def\urlprefix{URL }\fi
\expandafter\ifx\csname href\endcsname\relax
  \def\href#1#2{#2} \def\path#1{#1}\fi

\bibitem{Niknam2012-kl}
T.~Niknam, F.~Golestaneh, A.~Malekpour, Probabilistic energy and operation
  management of a microgrid containing wind/photovoltaic/fuel cell generation
  and energy storage devices based on point estimate method and self-adaptive
  gravitational search algorithm, Energy 43~(1) (2012) 427--437.

\bibitem{Neves2016-ga}
D.~Neves, M.~C. Brito, C.~A. Silva, Impact of solar and wind forecast
  uncertainties on demand response of isolated microgrids, Renewable Energy 87
  (2016) 1003--1015.

\bibitem{Iea2019-xm}
U.~Iea, Global status report for buildings and construction 2019, IEA Paris,
  France, 2019.

\bibitem{Center2020-hq}
B.~P. Center, Annual energy outlook 2020, Energy Information Administration,
  Washington, DC 12 (2020) 1672--1679.

\bibitem{Eckman2022-qo}
T.~Eckman, L.~C. Schwartz, G.~Leventis, Determining utility system value of
  demand flexibility from grid-interactive efficient buildings (Dec. 2022).

\bibitem{Hu2021-yw}
M.~Hu, F.~Xiao, S.~Wang, Neighborhood-level coordination and negotiation
  techniques for managing demand-side flexibility in residential microgrids,
  Renewable Sustainable Energy Rev. 135 (2021) 110248.

\bibitem{Li2021-ox}
H.~Li, Z.~Wang, T.~Hong, M.~A. Piette, Energy flexibility of residential
  buildings: A systematic review of characterization and quantification methods
  and applications, Advances in Applied Energy 3 (2021) 100054.

\bibitem{Li2022-gs}
R.~Li, A.~J. Satchwell, D.~Finn, T.~H. Christensen, M.~Kummert,
  J.~Le~Dr{\'e}au, R.~A. Lopes, H.~Madsen, J.~Salom, G.~Henze, K.~Wittchen, Ten
  questions concerning energy flexibility in buildings, Build. Environ. 223
  (2022) 109461.

\bibitem{Wang2019-kt}
H.~Wang, S.~Wang, R.~Tang, Development of grid-responsive buildings:
  Opportunities, challenges, capabilities and applications of {HVAC} systems in
  non-residential buildings in providing ancillary services by fast demand
  responses to smart grids, Appl. Energy 250 (2019) 697--712.

\bibitem{Watson2006-cp}
D.~S. Watson, S.~Kiliccote, N.~Motegi, M.~A. Piette, Strategies for demand
  response in commercial buildings,
  \url{https://escholarship.org/content/qt22t2d863/qt22t2d863.pdf}, accessed:
  2023-1-11 (2006).

\bibitem{Li2012-lu}
N.~Li, G.~Calis, B.~Becerik-Gerber, Measuring and monitoring occupancy with an
  {RFID} based system for demand-driven {HVAC} operations, Autom. Constr. 24
  (2012) 89--99.

\bibitem{Tang2018-jw}
R.~Tang, S.~Wang, C.~Yan, A direct load control strategy of centralized
  air-conditioning systems for building fast demand response to urgent requests
  of smart grids, Autom. Constr. 87 (2018) 74--83.

\bibitem{Wang2020-di}
H.~Wang, S.~Wang, K.~Shan, Experimental study on the dynamics, quality and
  impacts of using variable-speed pumps in buildings for frequency regulation
  of smart power grids, Energy (2020).

\bibitem{De_Chalendar2023-zh}
J.~A. de~Chalendar, C.~McMahon, L.~Fuentes~Valenzuela, P.~W. Glynn, S.~M.
  Benson, Unlocking demand response in commercial buildings: Empirical response
  of commercial buildings to daily cooling set point adjustments, Energy Build.
  278~(112599) (2023) 112599.

\bibitem{Yin2016-iy}
R.~Yin, E.~C. Kara, Y.~Li, N.~DeForest, K.~Wang, T.~Yong, M.~Stadler,
  Quantifying flexibility of commercial and residential loads for demand
  response using setpoint changes, Appl. Energy 177 (2016) 149--164.

\bibitem{Crawley2000-ji}
D.~B. Crawley, L.~K. Lawrie, C.~O. Pedersen, F.~C. Winkelmann, Energy plus:
  energy simulation program, ASHRAE J. 42~(4) (2000) 49--56.

\bibitem{Hu2017-tz}
M.~Hu, F.~Xiao, L.~Wang, Investigation of demand response potentials of
  residential air conditioners in smart grids using grey-box room thermal
  model, Appl. Energy 207 (2017) 324--335.

\bibitem{Hu2018-dg}
M.~Hu, F.~Xiao, Price-responsive model-based optimal demand response control of
  inverter air conditioners using genetic algorithm, Appl. Energy (2018).

\bibitem{Kiliccote2010-re}
S.~Kiliccote, M.~A. Piette, J.~Mathieu, K.~Parrish, Findings from seven years
  of field performance data for automated demand response in commercial
  buildings, Tech. Rep. LBNL-3643E, Lawrence Berkeley National Lab. (LBNL),
  Berkeley, CA (United States) (May 2010).

\bibitem{Aghniaey2018-ms}
S.~Aghniaey, T.~M. Lawrence, The impact of increased cooling setpoint
  temperature during demand response events on occupant thermal comfort in
  commercial buildings: A review, Energy Build. 173 (2018) 19--27.

\bibitem{Tremblay2022-wc}
C.~Tremblay, pyhaystack, \url{https://github.com/ChristianTremblay/pyhaystack}
  (2022).

\bibitem{Project_Haystack2022-ti}
{Project Haystack}, Project haystack,
  \url{https://www.project-haystack.org/about} (2022).

\bibitem{Brelih2012-pq}
N.~Brelih, How to improve energy efficiency of fans for air handling units,
  REHVA J 49 (2012) 5--10.

\bibitem{Tang2018-ch}
R.~Tang, S.~Wang, K.~Shan, Optimal and near-optimal indoor temperature and
  humidity controls for direct load control and proactive building demand
  response towards smart grids, Autom. Constr. 96 (2018) 250--261.

\bibitem{Ma2012-rc}
Y.~Ma, A.~Kelman, A.~Daly, F.~Borrelli, Predictive control for energy efficient
  buildings with thermal storage: Modeling, stimulation, and experiments, IEEE
  Control Syst. Mag. 32~(1) (2012) 44--64.

\bibitem{Zhuang2021-dl}
C.~Zhuang, K.~Shan, S.~Wang, Coordinated demand-controlled ventilation strategy
  for energy-efficient operation in multi-zone cleanroom air-conditioning
  systems, Build. Environ. 191 (2021) 107588.

\bibitem{Zhou2016-wy}
X.~Zhou, D.~Yan, Y.~Jiang, X.~Shi, Influence of asynchronous demand behavior on
  overcooling in multiple zone {AC} systems, Build. Environ. 110 (2016) 65--75.

\bibitem{Fan2017-pi}
C.~Fan, F.~Xiao, Y.~Zhao, A short-term building cooling load prediction method
  using deep learning algorithms, Appl. Energy 195 (2017) 222--233.

\bibitem{Reddy2007-mo}
T.~A. Reddy, I.~Maor, C.~Panjapornpon, Calibrating detailed building energy
  simulation programs with measured {Data---Part} {II}: Application to three
  case study office buildings ({RP-1051}), HVAC\&R Research 13~(2) (2007)
  243--265.

\bibitem{Balaji2013-tu}
B.~Balaji, H.~Teraoka, R.~Gupta, Y.~Agarwal, {ZonePAC}: Zonal power estimation
  and control via {HVAC} metering and occupant feedback, in: Proceedings of the
  5th {ACM} Workshop on Embedded Systems For {Energy-Efficient} Buildings,
  BuildSys'13, Association for Computing Machinery, New York, NY, USA, 2013,
  pp. 1--8.

\bibitem{MacDonald2020-ri}
J.~S. MacDonald, E.~Vrettos, D.~S. Callaway, A critical exploration of the
  efficiency impacts of demand response from {HVAC} in commercial buildings,
  Proc. IEEE 108~(9) (2020) 1623--1639.

\bibitem{Keskar2022-nj}
A.~Keskar, S.~Lei, T.~Webb, S.~Nagy, I.~A. Hiskens, J.~L. Mathieu, J.~X.
  Johnson, Assessing the performance of global thermostat adjustment in
  commercial buildings for load shifting demand response, Environ. Res.:
  Infrastruct. Sustain. 2~(1) (2022) 015003.

\bibitem{Beil2015-ip}
I.~Beil, I.~A. Hiskens, S.~Backhaus, Round-trip efficiency of fast demand
  response in a large commercial air conditioner, Energy Build. 97 (2015)
  47--55.

\bibitem{Afshari2017-qj}
S.~Afshari, J.~Wolfe, M.~S. Nazir, I.~A. Hiskens, J.~X. Johnson, J.~L. Mathieu,
  Y.~Lin, A.~K. Barnes, D.~A. Geller, S.~N. Backhaus, An experimental study of
  energy consumption in buildings providing ancillary services, in: 2017 {IEEE}
  Power \& Energy Society Innovative Smart Grid Technologies Conference
  ({ISGT}), ieeexplore.ieee.org, 2017, pp. 1--5.

\bibitem{Raftery2018-xe}
P.~Raftery, S.~Li, B.~Jin, M.~Ting, G.~Paliaga, H.~Cheng, Evaluation of a
  cost-responsive supply air temperature reset strategy in an office building,
  Energy Build. 158 (2018) 356--370.

\end{thebibliography}

\newpage
\appendix

\section{Supplemental tables and figures}
\label{sec: appendix}

\begin{figure}[htbp]
    \centering
    \includegraphics[width=.9\textwidth]{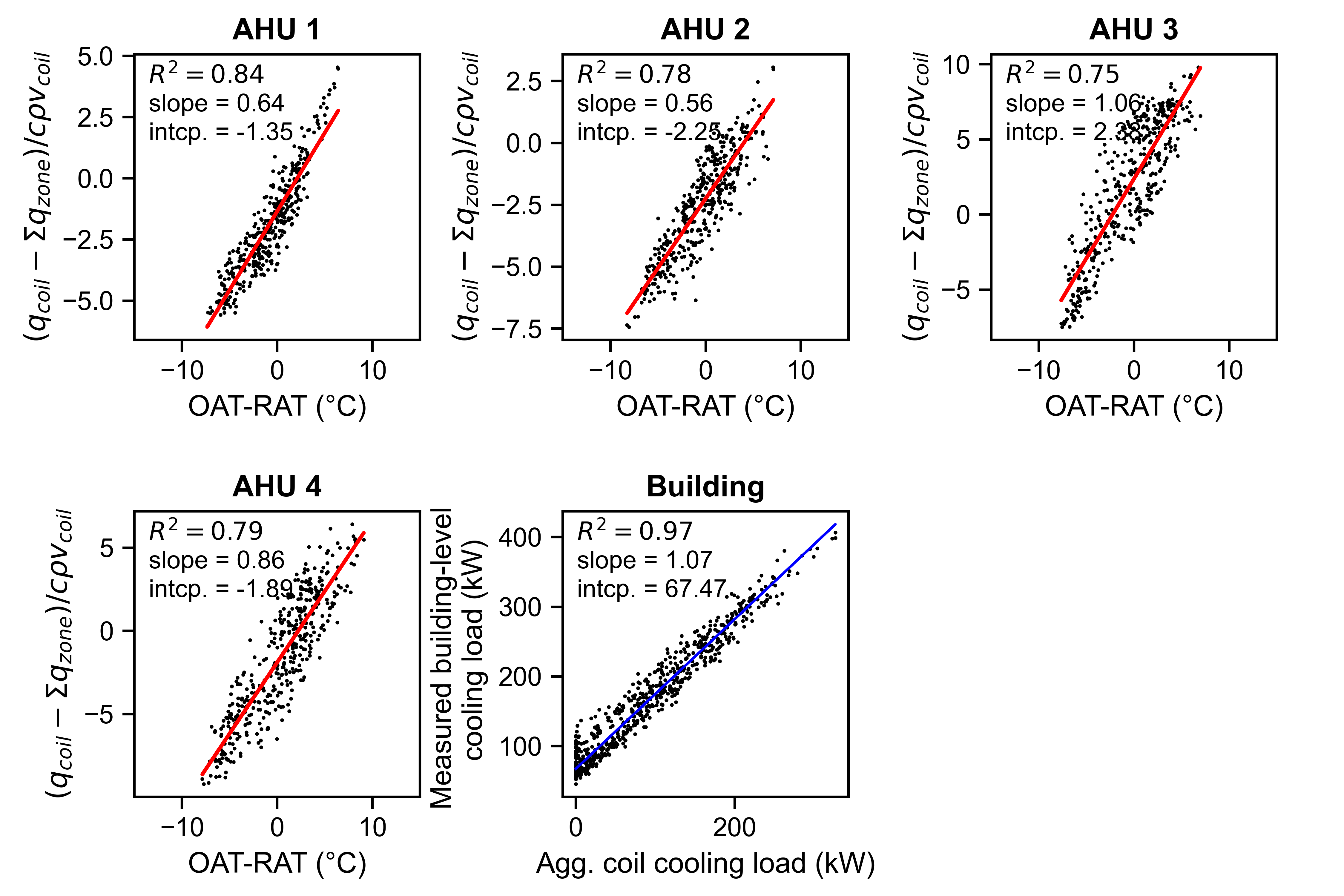}
    \caption{Identification results for Bldg. B during training session, including regression models from aggregate zone-level space cooling load to AHU coil cooling load for each AHU and regression model from aggregate coil cooling load to measured building cooling load}
    \label{fig:appendix-1}
\end{figure}

\begin{figure}[htbp]
    \centering
    \includegraphics[width=.9\textwidth]{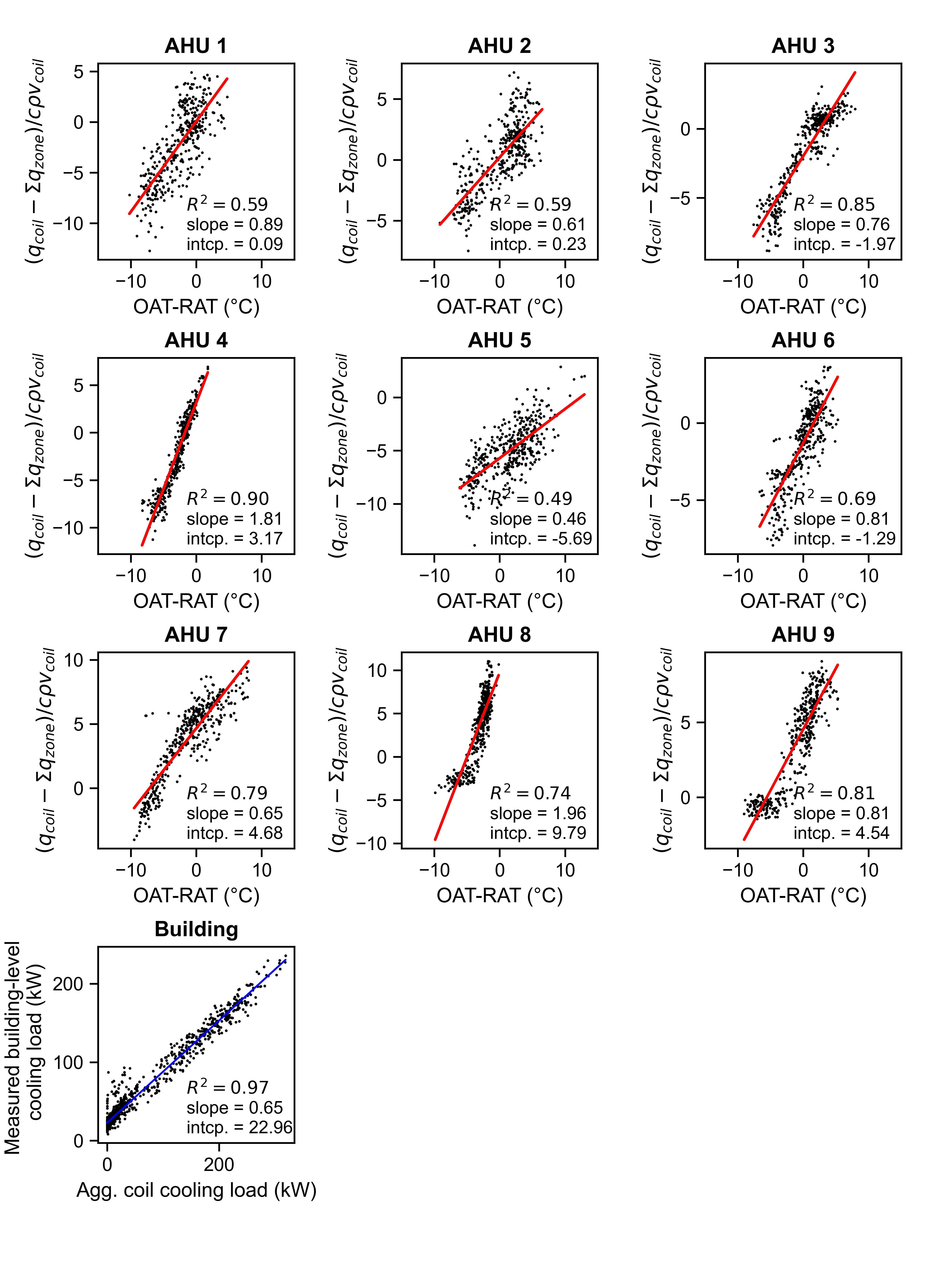}
    \caption{Identification results for Bldg. C during training session, including regression models from aggregate zone-level space cooling load to AHU coil cooling load for each AHU and regression model from aggregate coil cooling load to measured building cooling load}
    \label{fig:appendix-2}
\end{figure}

\begin{table}[ht]
    \centering
    \caption{
    \label{tab: regression models for building A} Additional numerical data for the models shown in \Cref{fig:regression models_bldg_A} (Bldg. A, Training session)}
    \begin{tabular}{lllllll}
    \hline
     & AHU 1 & AHU 2 & AHU 3 & AHU 4 & AHU 5 & Building \\
    \hline
    Number of observations & 380  &  386 & 385 & 385  &  386 & 1032 \\
    $R^2$ & 0.926 & 0.800 & 0.761 & 0.806 & 0.761 & 0.918 \\
    Prob (F-statistic) & 2.82e-216 & 1.97e-136 & 3.46e-121 & 1.46e-138 & 1.46e-121 & 0.00 \\
    \hline 
    \end{tabular}%

    \begin{tabular}{c|ccc|ccc|ccc}
    \hline 
     &  \multicolumn{3}{c|}{AHU 1} & \multicolumn{3}{c|}{AHU 2} & \multicolumn{3}{c}{AHU 3}\\
     \hline
     & Value & Std. Err & P-Value & Value & Std. Err & P-Value & Value & Std. Err & P-Value \\
    \hline
    Intercept & -0.542  &  0.029 & 0.000 & -0.430 &  0.046 & 0.000 & -1.232  & 0.041 & 0.000  \\
    Slope & 0.512 & 0.007 & 0.000 & 0.428 & 0.011 & 0.000 & 0.488  & 0.014 & 0.000  \\
    \hline
    \end{tabular}%

    \begin{tabular}{c|ccc|ccc|ccc}
    \hline 
     &  \multicolumn{3}{c|}{AHU 4} & \multicolumn{3}{c|}{AHU 5} & \multicolumn{3}{c}{Building}\\
     \hline
     & Value & Std. Err & P-Value & Value & Std. Err & P-Value & Value & Std. Err & P-Value \\
    \hline
    Intercept & -0.371  &  0.049 & 0.000 & -0.219  &  0.057 & 0.000 & 29.802  &  1.148 & 0.000  \\
    Slope & 0.781 & 0.020 & 0.000 & 0.495 & 0.014 & 0.000 & 1.259  &  0.012 & 0.000  \\
    \hline
    \end{tabular}%
\end{table}

\begin{table}[ht]
    \centering
    \caption{
    \label{tab: regression models for building B} Additional numerical data for the models shown in \Cref{fig:appendix-1} (Bldg. B, Training session)}
    \begin{tabular}{llllll}
    \hline
     & AHU 1 & AHU 2 & AHU 3 & AHU 4 & Building \\
    \hline
    Number of observations & 403 & 403 & 403 & 403  & 1032 \\
    $R^2$ & 0.843 & 0.776 & 0.750 & 0.788 & 0.966 \\
    Prob (F-statistic) & 1.43e-163 & 3.05e-132 & 6.85e-123 & 3.90e-137 & 0.00 \\
    \hline 
    \end{tabular}%

    \begin{tabular}{c|ccc|ccc|ccc}
    \hline 
     &  \multicolumn{3}{c|}{AHU 1} & \multicolumn{3}{c|}{AHU 2} & \multicolumn{3}{c}{AHU 3}\\
     \hline
     & Value & Std. Err & P-Value & Value & Std. Err & P-Value & Value & Std. Err & P-Value \\
    \hline
    Intercept & -1.348  &  0.042 & 0.000 & -2.249 &  0.049 & 0.000 & 2.378  & 0.110 & 0.000  \\
    Slope & 0.642 & 0.014 & 0.000 & 0.561 & 0.015 & 0.000 & 1.058  & 0.030 & 0.000  \\
    \hline
    \end{tabular}%

    \begin{tabular}{c|ccc|ccc}
    \hline 
     & \multicolumn{3}{c|}{AHU 4} & \multicolumn{3}{c}{Building}\\
     \hline
     & Value & Std. Err & P-Value & Value & Std. Err & P-Value \\
    \hline
    Intercept & -1.893  &  0.084 & 0.000 & 67.475  &  0.609 & 0.000 \\
    Slope & 0.856 & 0.022 & 0.000 & 1.073 & 0.006 & 0.000 \\
    \hline
    \end{tabular}%
\end{table}

\begin{table}[ht]
    \centering
    \caption{
    \label{tab: regression models for building C} Additional numerical data for the models shown in \Cref{fig:appendix-2} (Bldg. C, Training session)}
    \begin{tabular}{llllll}
    \hline
     & AHU 1 & AHU 2 & AHU 3 & AHU 4 & AHU 5 \\
    \hline
    Number of observations & 390  &  390 & 390 & 390  &  403 \\
    $R^2$ & 0.587 & 0.591 & 0.845 & 0.899 & 0.493  \\
    Prob (F-statistic) & 1.34e-76 & 2.54e-77 & 3.65e-159 & 1.94e-195 & 4.66e-61 \\
    \hline 
    \end{tabular}%
    
    \begin{tabular}{llllll}
    \hline
     & AHU 6 & AHU 7 & AHU 8 & AHU 9 & Building \\
    \hline
    Number of observations & 403 & 403 & 403 & 403 & 1032 \\
    $R^2$ & 0.689 & 0.785 & 0.742 & 0.813 & 0.971  \\
    Prob (F-statistic) & 1.26e-103 & 4.30e-136 & 4.45e-120 & 3.16e-148 & 0.00 \\
    \hline 
    \end{tabular}%

    \begin{tabular}{c|ccc|ccc|ccc}
    \hline 
     &  \multicolumn{3}{c|}{AHU 1} & \multicolumn{3}{c|}{AHU 2} & \multicolumn{3}{c}{AHU 3}\\
     \hline
     & Value & Std. Err & P-Value & Value & Std. Err & P-Value & Value & Std. Err & P-Value \\
    \hline
    Intercept & 0.090  &  0.150 & 0.546 & 0.232 &  0.090 & 0.010 & -1.967  & 0.060 & 0.000  \\
    Slope & 0.893 & 0.038 & 0.000 & 0.606 & 0.026 & 0.000 & 0.763  & 0.017 & 0.000  \\
    \hline
    \end{tabular}%

    \begin{tabular}{c|ccc|ccc|ccc}
    \hline 
     &  \multicolumn{3}{c|}{AHU 4} & \multicolumn{3}{c|}{AHU 5} & \multicolumn{3}{c}{AHU 6}\\
     \hline
     & Value & Std. Err & P-Value & Value & Std. Err & P-Value & Value & Std. Err & P-Value \\
    \hline
    Intercept & 3.173  &  0.106 & 0.000 & -5.694  &  0.093 & 0.000 & -1.293  &  0.069 & 0.000  \\
    Slope & 1.808 & 0.031 & 0.000 & 0.462 & 0.023 & 0.000 & 0.814  &  0.027 & 0.000  \\
    \hline
    \end{tabular}%
    
    \begin{tabular}{c|ccc|ccc|ccc}
    \hline 
     &  \multicolumn{3}{c|}{AHU 7} & \multicolumn{3}{c|}{AHU 8} & \multicolumn{3}{c}{AHU 9} \\
     \hline
     & Value & Std. Err & P-Value & Value & Std. Err & P-Value & Value & Std. Err & P-Value \\
    \hline
    Intercept & 4.683  &  0.072 & 0.000 & 9.789  &  0.212 & 0.000 & 4.542  &  0.071 & 0.000  \\
    Slope & 0.654 & 0.017 & 0.000 & 1.958 & 0.058 & 0.000 & 0.814  &  0.019 & 0.000  \\
    \hline
    \end{tabular}%

    \begin{tabular}{c|ccc}
    \hline 
     &  \multicolumn{3}{c}{Building}\\
     \hline
     & Value & Std. Err & P-Value \\
    \hline
    Intercept & 22.959 & 0.371 & 0.000 \\
    Slope & 0.651 & 0.004 & 0.000  \\
    \hline
    \end{tabular}%
\end{table}

\begin{table}[ht]
    \centering
    \caption{
    \label{tab: comparison between training and validation sessions} Performance of regression models ($R^2$) during training and validation sessions for each target building}
    \begin{tabular}{lllllll}
    \hline
     & AHU 1 & AHU 2 & AHU 3 & AHU 4 & AHU 5 & Building \\
    \hline
    \# of observations (Bldg. A, Training) & 380 & 386 & 385 & 385 & 386 & 1032 \\
    $R^2$ (Bldg. A, Training) & 0.926  &  0.800 & 0.761 & 0.806  &  0.761 & 0.918 \\
    \# of observations (Bldg. A, Validation) & 95 & 97 &96 & 96 & 96 & 264 \\
    $R^2$ (Bldg. A, Validation) & 0.915 & 0.554 & 0.588 & 0.675 & 0.692 & 0.840  \\
    \hline 
    \end{tabular}%
    
    \begin{tabular}{llllll}
    \hline
     & AHU 1 & AHU 2 & AHU 3 & AHU 4 & Building \\
    \hline
    \# of observations (Bldg. B, Training) & 403 & 403 & 403 & 403 & 1032 \\
    $R^2$ (Bldg. B, Training) & 0.843  &  0.776 & 0.750 & 0.788  &  0.966 \\
    \# of observations (Bldg. B, Validation) & 104 & 104 & 104 & 104 & 264 \\
    $R^2$ (Bldg. B, Validation) & 0.835 & 0.675 & 0.831 & 0.835 & 0.865  \\
    \hline 
    \end{tabular}%

    \begin{tabular}{lllllll}
    \hline
     & AHU 1 & AHU 2 & AHU 3 & AHU 4 & AHU 5& AHU 6 \\
    \hline
    \# of observations (Bldg. C, Training) & 390 & 390 & 390 & 390 & 403 & 403 \\
    $R^2$ (Bldg. C, Training) & 0.587  &  0.591 & 0.845 & 0.899  &  0.493 & 0.689 \\
    \# of observations (Bldg. C, Validation) & 104 & 104 & 104 & 104 & 104 & 104 \\
    $R^2$ (Bldg. C, Validation) & 0.311 & 0.743 & 0.829 & 0.934 & 0.487 & 0.751  \\
    \hline 
    \end{tabular}%

    \begin{tabular}{lllll}
    \hline
     & AHU 7 & AHU 8 & AHU 9 & Building\\
    \hline
    \# of observations (Bldg. C, Training) & 403 & 403 & 403 & 1032 \\
    $R^2$ (Bldg. C, Training) & 0.785  &  0.742 & 0.813 & 0.971\\
    \# of observations (Bldg. C, Validation) & 104 & 104 & 104 & 264 \\
    $R^2$ (Bldg. C, Validation) & 0.872 & 0.707 & 0.862 & 0.881\\
    \hline 
    \end{tabular}%

\end{table}

\end{document}